\documentclass[preprint,tighten,a4paper,showpacs]{revtex4}
\usepackage{epsfig,graphicx}
\usepackage{amsfonts}
\usepackage{amsbsy}
\newcommand{\beq}{\begin{equation}}
\newcommand{\eeq}[1]{\label{#1}\end{equation}}

\newcommand{\be}{\begin{eqnarray}}
\newcommand{\ee}{\end{eqnarray}}

\newcommand{\vecsig}{\vec\sigma}
\newcommand{\vecq}{\vec q}
\newcommand{\pwave}{$p$-wave}

\begin{document}

\pagestyle{plain}
\baselineskip 16pt\vskip 48pt

\title{
Chiral approach to antikaon \mbox{\boldmath \it s}- and 
\mbox{\boldmath \it p}-wave interactions
in dense nuclear matter}

\author{L. Tol\'os}
\affiliation{
   Gesellschaft f\"ur Schwerionenforschung,
   Planckstrasse 1,
   D-64291 Darmstadt, Germany
}

\author{A. Ramos}

\affiliation{
     Departament d'Estructura i Constituents de la Mat\`{e}ria,
     Universitat de Barcelona,
     Diagonal 647, 08028 Barcelona, Spain
}

\author{E. Oset}

\affiliation{
     Departamento de F\'{\i}sica Te\'orica and IFIC, 
     Centro Mixto Universidad de Valencia-CSIC,
     Institutos de Investigaci\'on de Paterna, 
     Ap. Correos 22085, E-46071 Valencia, Spain \\ }

\begin{abstract}
The properties of the antikaons in nuclear matter are investigated from a
chiral unitary approach which incorporates the $s$- and $p$-waves of the ${\bar
K}N$ interaction. 
To obtain the in-medium meson-baryon amplitudes we include, in a
self-consistent way, Pauli blocking effects, meson self-energies corrected 
by nuclear short-range correlations and baryon binding potentials.
We pay  special attention to investigating the validity of the on-shell
factorization, showing that it
cannot be applied in the evaluation of the in-medium corrections to the 
$p$-wave amplitudes. In nuclear matter at saturation energy, 
the $\Lambda$ and $\Sigma$ develop an attractive potential of about -30
MeV, while the $\Sigma^*$ pole remains at the free space value although
its width gets sensibly increased to about 80 MeV. 
The antikaon also develops a moderate attraction that does not support the
existence of very deep and narrow bound states, confirming the findings of previous
self-consistent calculations.

\end{abstract}

\date{\today}

\maketitle

\section{Introduction} 
The interaction of $\bar{K}$ with nucleons
and nuclei has captured much interest in recent times \cite{review}. The
elementary interaction close to threshold is linked to the presence below
threshold of the $\Lambda(1405)$ resonance, long claimed to be a dynamically
generated resonance in coupled channels \cite{dalitz, jennings} more than a
genuine three constituent quark state.  The advent of chiral theories 
\cite{Gasser:1984gg,Meissner:1993ah,Bernard:1995dp,Pich:1995bw,Ecker:1994gg} and
unitarization extensions in coupled channels has brought a new perspective into
the topic and reconfirmed these first claims
\cite{kaiser,Oset:1997it,Oller:2000fj,lutz02,hyodo}.  It has also allowed to tackle the
problem in a more systematic way and has brought some surprises, as the
existence of more dynamically generated resonances like the $\Lambda(1670)$
\cite{Oset:2001cn,Garcia-Recio:2002td,cola}, the $\Xi(1620)$ \cite{bennhold2}, the $\Xi(1690)$
\cite{carmen},
hints of some resonance in $I=1$ and $S=-1$ in \cite{Oller:2000fj}, etc.  A thorough
work, looking at the SU(3) group structure of the generated states, has allowed a
generalization of the problem concluding that there is a singlet and two octets
of states which are dynamically generated in the interaction of the octet of
pseudoscalar mesons with the octet of $1/2^+$ baryons \cite{cola,carmen}. There is
also a surprise in this investigation which concludes that the actual
experimental $\Lambda(1405)$ is a superposition of two states and that this
should have experimental repercussions by making the $\Lambda(1405)$ appear with
different energy and width in different reactions.  This claim had a recent
experimental confirmation in the experiment of \cite{prakhov} and the posterior
analysis of the experiment in \cite{magas}.

    The interaction of $\bar{K}$ in nuclei has had a parallel and equally
exciting development. The $K^- N$ scattering matrix at threshold is repulsive.
However, the phenomenology of kaonic atoms demands an attractive potential
\cite{gal,baca}, which is telling us that, in spite of the  small nuclear
densities experienced by the $K^-$ atoms, the low density theorem $\Pi = t
\rho$, where $\Pi$ is the $K^-$ self-energy ($2 \omega V_{\rm opt}$), $t$ the
scattering matrix and $\rho$ the nuclear density, breaks down already
 at these small densities.  The presence of the $\Lambda(1405)$ resonance just below
$\bar{K}N$ threshold is at the root of this problem.  The first step to
understand this change of sign was given in \cite{koch} (see also \cite{waas})
which showed that the consideration of Pauli blocking in the intermediate
$\bar{K} N$ states moved the position of the $\Lambda(1405)$ resonance, or,
equivalently, the zero of the real part of the ${\bar K}N$ amplitude, to higher
energies, hence passing from a free-space repulsive
scattering matrix at threshold to an attractive one in the medium.
The  problem is more
subtle, as was shown in \cite{lutz} (see also \cite{schaffner}), since the
self-consistent consideration of the generated attractive self-energy in the
$\bar{K}$ has as a consequence the shift of the resonance position to lower 
energies introducing
a repulsion.  The final self-consistent solution still leads to a moderate
attraction on the $\bar{K}$, but much smaller than by just considering the Pauli
blocking effect.  A further consideration of the self-energies of the pions and baryons
in the intermediate states was done in \cite{Ramos:1999ku} opening new decay
channels for the $\bar{K}$ but not modifying much the real part of the
potential.  Moderate attractions of the order of -50 MeV at full nuclear matter
density are found in these approaches and the potential obtained is shown to
fairly reproduce the data on kaonic atoms \cite{okumura}. This fairness is
further investigated in \cite{baca} where a fit to data is made in order to see
how far is the calculated potential from an optimal one. The potential of 
\cite{Ramos:1999ku} also leads to deeply bound $K^-$ states in medium and heavy
nuclei which are bound by about 30-40 MeV but have a width of about 100 MeV 
\cite{okumura}.

   Different steps in this problem were given in \cite{akaishi}, constructing a
kaon-nucleus potential with very large strength and predicting strongly bound
states in few body systems.  This phenomenological potential has been
critically discussed in \cite{toki}, where it is shown that the omission of the
direct coupling of the $\pi \Sigma$ channel to itself, the assumption of the
nominal $\Lambda(1405)$ as a single bound $\bar{K}$ state --while there are 
two states in the
chiral unitary approach--, the lack of self-consistency in the calculations and
the seemingly too large nuclear densities obtained of around ten times normal nuclear
matter density at the center of the nucleus, lead altogether to a much larger
potential than that obtained in the chiral approaches.  With the potential
of    \cite{akaishi} a bound state of $K^-$ in $^3$He by 108 MeV was predicted
in $I=0$.  Later on an experiment at KEK found a peak in the proton spectrum
following  the absorption of stopped $K^-$ by $^4$He \cite{suzuki}, which was
interpreted as a strange tribaryon with $I=1$, and not as a kaon atom, since its
interpretation as a kaon atom required 195 MeV binding and the isospin was also
different than the predictions made in \cite{akaishi}.  However, further work
done in \cite{akaishi2} including relativistic corrections (always considered
in the chiral approaches), spin orbit effects and some further ad hoc
changes produced a binding in the three body system compatible with the
association of the tribaryon state claimed in \cite{suzuki} to the kaon
bound state. At this point the $K^-$-nucleus potential has a strength of 615
MeV at the center of the nucleus, roughly twenty times what one would get from the
chiral theories for this problem.  Such disparate approaches led the authors of
\cite{toki} to search for a different interpretation of the peaks found in the
experiment of \cite{suzuki}, and an explanation of the peaks was found based on
the mechanism of $K^-$ absorption in nucleon pairs leading to $\Lambda p$ and
$\Sigma p$ without further interaction of the final pair of particles with the
nucleus. 

    Another experiment, looking at the $\Lambda p$ invariant mass after $K^-$
absorption in different light and medium nuclei, interpreted a broad peak as a
bound state of $K^- pp$ by 115 MeV \cite{FINUDA}.  Such a large number and the 
alternative
explanation found for the KEK peaks, motivated the authors of \cite{angelsmagas}
to look for another interpretation of the peak, which was found in the process
of $K^-$ absorption by pairs of nucleons leading to $\Lambda N$, followed by the
scattering of the $\Lambda$ or the  $N$ with the daughter nucleus. 

   On the theoretical side there have been further advances by looking at the
$p$-wave contribution to the optical potential \cite{angelscarmen,laura,lutz-korpa02}. 
 In \cite{angelscarmen} the
$p$-wave $K^-$-nucleus optical potential was studied for atoms and found to be
basically negligible. Yet, for other situations where the momenta of the kaon
could be larger than in the atoms, this contribution could be bigger and it is
thus mandatory to address this problem. Some work in this direction has already
been done in \cite{laura}, where the momentum dependence of the antikaon 
potential in nuclear matter was obtained from a self-consistent calculation 
using the meson-exchange J\"ulich interaction. At momenta as large as 
500 MeV/c, the higher partial waves beyond $L=0$ modify strongly the 
$\bar K$ potential. 

    In view of the interest and controversy of the subject, the importance of
having an accurate as possible description of the interaction of $\bar{K}$ with
nucleons in nuclear matter is rather evident and it is our purpose to present
here the results of a careful study of the $s$- and $p$-wave contributions to
the $\bar{K}$ nucleus potential.  The work follows closely the lines of
\cite{Ramos:1999ku}, incorporating the many-body corrections to the $p$-wave
$\bar{K} N$ interaction studied in \cite{Jido:2002zk} within the same chiral
unitary approach. An evaluation of the antikaon propagation in matter based on
chiral dynamics including $s$-, $p$- and $d$-waves was already performed in
\cite{lutz-korpa02}. We will investigate the validity of the on-shell
factorization of the kernel in the Bethe-Salpeter equation. This approach is
often used in the study of the hadron-hadron collisions and finds its
justification in the N/D dispersion relation method used in \cite{Oller:2000fj}
and \cite{Oller:1998zr}.  However, this is not justified when one goes to the
nuclear medium since there are new sources of imaginary part different from
those of the free case. As we shall see, the use of this prescription in the
nuclear medium  for $p$-waves leads to a violation of causality, producing
negative decay widths  in some cases. A different sort of on-shell
aproximation is done in Ref. \cite{lutz-korpa02}, which in practice leads to
quite diferent results for $p$-waves than those reported here.

   A byproduct of any calculation including a self-consistent propagation of
   $p$-waves in a coupled-channel scheme is the self-energy of the 
   hyperons involved in the $p$-wave ${\bar K}N$ interaction kernel.
   At variance to Ref.~\cite{lutz-korpa02}, the in-medium ${\bar K}N$
   amplitudes calculated in the present work
   include self-energy insertions not only on the antikaons but
   also on the pions present as intermediate states in the coupled-channel
   scheme. This gives rise to a more realistic derivation of the hyperon
   self-energies, as we shall see, since important
   $YN$ interaction pieces, such as $\Lambda N \to \Sigma N$, conveniently
   modified by the effect of short-range correlations are also included in
   the present work.
   
   The paper is organized as follows. We first recall in Sect.~\ref{free} 
   the structure of the
   $s$- and $p$-wave ${\bar K}N$ amplitudes in free space. Next, in
   Sect.~\ref{medium}, we describe how the different medium effects 
   are incorporated into our scheme, paying special
   attention to discussing the validity of the on-shell
   factorization in the medium as well as to showing how the intermediate meson
   propagators get modified by the effect of short-range correlations. In
   Sect.~\ref{results} we present and discuss our
   results for the in-medium amplitudes and the antikaon properties, such as the
   spectral function or the optical potential. Finally, our concluding remarks
   are given in Sect.~\ref{conclusions}.

\section{Meson-baryon amplitudes in free space}
\label{free}

The lowest order chiral Lagrangian which couples the octet of pseudoscalar mesons to the
octet of $1/2^+$ baryons is given by \cite{Meissner:1993ah,Bernard:1995dp,Pich:1995bw,Ecker:1994gg} 
\begin{eqnarray}
   {\cal L}_1^{(B)} &=& \langle \bar{B} i \gamma^{\mu} \nabla_{\mu} B
    \rangle  - M \langle \bar{B} B\rangle  \nonumber \\
    &&  + \frac{1}{2} D \left\langle \bar{B} \gamma^{\mu} \gamma_5 \left\{
     u_{\mu}, B \right\} \right\rangle + \frac{1}{2} F \left\langle \bar{B}
     \gamma^{\mu} \gamma_5 \left[u_{\mu}, B\right] \right\rangle  \ ,
    \label{chiralLag}
\end{eqnarray}
where $B$ is the SU(3) matrix for baryons, $M$ is the baryon mass, $u$ contains the $\Phi$ matrix of mesons and the symbol $\langle \, \rangle$ denotes the trace of SU(3) flavor
matrices. The SU(3) matrices appearing in Eq.~(\ref{chiralLag}) are standard and can be seen in the former references. The couplings $D$ and $F$ are chosen as $D=0.85$ and $F=0.52$.

At lowest order, the $BB\Phi\Phi$ interaction Lagrangian reads
\begin{equation}
   {\cal L}_1^{(B)} = \left\langle \bar{B} i \gamma^{\mu} \frac{1}{4 f^2}
   [(\Phi\, \partial_{\mu} \Phi - \partial_{\mu} \Phi \Phi) B
   - B (\Phi\, \partial_{\mu} \Phi - \partial_{\mu} \Phi \Phi)] 
   \right\rangle \ . \label{lowest}
\end{equation}
For low energies, one derives the $s$-wave amplitude
\begin{eqnarray}
V_{i j}^s= - C_{i j} \, \frac{1}{4 f^2} \, (2 \, \sqrt{s}-M_{B_i}-M_{B_j}) 
\left( \frac{M_{B_i}+E_i}{2 \, M_{B_i}} \right)^{1/2} \, \left( \frac{M_{B_j}+E_j}{2 \, M_{B_j}} \right)^{1/2} \ ,
\label{swa}
\end{eqnarray}
being $M_{B_i}$ and $E_i$ the mass and energy of the baryon in the $i$ channel, respectively.
The coefficients $C_{ij}$ form a
symmetric matrix and are written explicitly in
\cite{Oset:1997it}. Following Ref.~\cite{Oset:1997it}, the meson decay constant $f$
is taken as an average value $f=1.123 f_\pi$ \cite{Oset:2001cn}. The
channels included in our study are $K^- p$, $\bar{K}^0n$, $\pi^0
\Lambda$, $\pi^0 \Sigma^0$, $\eta \Lambda$, $\eta \Sigma^0$, $\pi^+
\Sigma^-$, $\pi^- \Sigma^+$, $K^+ \Xi^-$, $K^0 \Xi^0$. We note, for later
purposes, that this equation can be written to a good approximation as
\begin{eqnarray}
V_{i j}^s \simeq - C_{i j} \, \frac{1}{4 f^2} \, (k^0_j + k^0_i) \ ,
\label{swa2}
\end{eqnarray}
where $k^0_i$ and $k^0_j$  are the initial and final meson
energies in the center-of-mass (c.m.) frame. 
The lagrangian in Eq.~(\ref{lowest}) provides also
a small part of the $p$-wave, which in the c.m. frame reads:
\begin{equation}
   V^{\, c}_{ij} = - C_{ij} {1 \over 4 f^2}\, a_i\, a_j \left({1 \over
   b_i} + {1 \over b_j} \right) (\vec\sigma \cdot \vec q_j)
   (\vec\sigma \cdot \vec q_i) \label{pwcont} \ ,
\end{equation}
with 
\begin{equation}
   a_i = \sqrt{E_i + M_i \over 2 M_i}\ , \hspace{0.7cm} b_i = E_i +
   M_i\ , \hspace{0.7cm} E_i = \sqrt{M_i^{\, 2} + \vec q_i\,^{ 2}} \ .
\end{equation}

The main contribution to the $p$-wave amplitude comes from the $\Lambda$ and $\Sigma$
pole terms which are obtained from the $D$ and $F$ terms of the
Lagrangian of Eq.~(\ref{chiralLag}) \cite{Jido:2002zk}. The $\Sigma^*$ pole term is also
included explicitly with couplings to the meson-baryon states
evaluated using SU(6) symmetry arguments \cite{Oset:2000eg}.
These contributions are given by:
\begin{eqnarray}
   V^{\Lambda}_{ij} &=& D^{\Lambda}_i D^{\Lambda}_j { 1 \over \sqrt{s}
   - \tilde M_\Lambda} (\vecsig \cdot \vecq_j)(\vecsig \cdot \vecq_i)
   \left(1+{q_j^0 \over M_j}\right) \left(1+{q_i^0 \over M_i} \right) 
   \nonumber \ , \\
   V^{\Sigma}_{ij} &=& D^{\Sigma}_i D^{\Sigma}_j { 1 \over \sqrt{s}
   - \tilde M_\Sigma} (\vecsig \cdot \vecq_j)(\vecsig \cdot \vecq_i)
   \left(1+{q_j^0 \over M_j} \right) \left(1+{q_i^0 \over M_i} \right) 
   \label{poleamps}\ , \\
   V^{\Sigma^*}_{ij} &=& D^{\Sigma^*}_i D^{\Sigma^*}_j { 1 \over \sqrt{s}
   - \tilde M_{\Sigma^*}} (\vec S \cdot \vecq_j)(\vec S^\dagger \cdot
   \vecq_i) \nonumber  \ ,
\end{eqnarray}
with $S^\dagger$ being the spin transition operator from spin $1/2$ to spin $3/2$
and
\begin{eqnarray}
   D^\Lambda_i &=& c_i^{D,\Lambda} \sqrt{20 \over 3} {D \over 2 f} -
   c_i^{F,\Lambda} \sqrt{12} { F \over 2 f} \nonumber \ , \\
   D^\Sigma_i &=& c_i^{D,\Sigma} \sqrt{20 \over 3} {D \over 2 f} -
   c_i^{F,\Sigma} \sqrt{12} { F \over 2 f} \ , \\
   D^{\Sigma^*}_i &=& c_i^{S,\Sigma^*} {12 \over 5} {D + F\over 2 f}
   \nonumber \ .
\end{eqnarray}
The constants $c^D$, $c^F$, $c^S$ are SU(3) Clebsch-Gordan
coefficients which depend upon the meson and baryon involved in the
vertex and are given in Table~I of Ref.~\cite{Jido:2002zk}. 
The masses $\tilde M_\Lambda$,
$\tilde M_\Sigma$, $\tilde M_{\Sigma^*}$ are bare masses of the
hyperons ($\tilde M_\Lambda$$=$1030 MeV, $\tilde M_\Sigma$$=$1120 MeV, $\tilde M_{\Sigma^*}$$=$1371 MeV),  which will turn into physical masses upon unitarization.

Once the tree level contributions to the $s$- and $p$-wave meson-baryon scattering are known, the Bethe-Salpeter equation can be solved using the tree level contributions as the kernel of the equation. In Ref.~\cite{Oset:1997it} it was shown that the kernel for the $s$-wave amplitude can
be factorized on the mass shell in the loop functions, by making some
approximations typical of heavy-baryon perturbation theory. Then the Bethe-Salpeter equation turns out to be simpler to solve. Furthermore, the factorization for {\pwave}s in
meson-meson scattering is also proved in \cite{Cabrera:2002hc} along the same
lines. A more general proof of the factorization is done
in \cite{Oller:1998zr} for meson-meson interactions and in
\cite{Oller:2000fj} for meson-baryon ones. 

The formal result obtained is schematically given by
\begin{equation}
    T = V + V G T \nonumber ,
\end{equation}
that is 
\begin{equation}
    T = [1-VG]^{-1} V \label{BSeq} \ , 
\end{equation}
where $V$ is the kernel (potential), given by the $s$- and $p$-wave amplitudes of 
Eqs.~(\ref{swa}),(\ref{pwcont}),(\ref{poleamps}), and $G$ is a diagonal matrix 
accounting for the loop function of a meson-baryon propagator, which needs to be
regularized. This can be done by adopting either a cut-off method or by using
dimensional regularization.
The cut-off method is easier and more transparent when dealing with particles
in the medium as it will be our case. The use of this cut-off scheme or the
dimensional regularization are in practice identical, given the matching
between the two loop functions done in Section 2 of Appendix {\it A} of
Ref.~\cite{pelaez}. There, one finds that the dimensional regularization
formula and the one with cut-off have the same analytical properties (the
log-terms) and are numerically equivalent for values of the cut-off reasonably
larger than the on-shell momentum of the states in the loop, which is a
condition respected in our calculations. By fine tuning the subtraction
constant in dimensional regularization, or fine tuning the cut-off, one can make
the two expressions identical at one energy and practically equal in a wide
range of energies, sufficient for studies like the present one. Using the
cut-off regularization the loop function reads in the c.m. frame

\begin{eqnarray} 
\hspace{-0.5cm}G_{l}(\sqrt{s})&=& i \, \int \frac{d^4 q}{(2
\pi)^4} \, \frac{M_l}{E_l (-\vec{q}\,)} \, \frac{1}{\sqrt{s} - q^0 - E_l
(-\vec{q}\,) + i \epsilon} \, \frac{1}{q^2 - m^2_l + i \epsilon} \nonumber \\
&=& \int_{\mid {\vec q}\, \mid < q_{\rm max}} \, \frac{d^3 q}{(2 \pi)^3} \,
\frac{1}{2 \omega_l (\vec q\,)} \frac{M_l}{E_l (-\vec{q}\,)} \,
\frac{1}{\sqrt{s}- \omega_l (\vec{q}\,) - E_l (-\vec{q}\,) + i \epsilon} \, ,
\label{eq:gprop} 
\end{eqnarray} 
with $\sqrt{s}=k^0+p^0$, being $p^0(k^0)$ the
energy of the initial baryon (meson) and $q_{\rm max}=630$~MeV.

The $p$-wave amplitudes of Eqs.~(\ref{pwcont}),(\ref{poleamps})
cannot be used directly in Eq.~(\ref{BSeq}) since, due to their spin 
structure, there is a mixture of different angular momenta.  As done in Ref.~\cite{Jido:2002zk},
we separate the $p$-wave amplitudes according to the total angular momentum. 
With the partial-wave amplitude written for $L=1$ as
\begin{eqnarray}
T(\vecq\,^\prime, \vecq\,) = (2L +1) \left( f(\sqrt{s})\, \hat
 q^{\prime} \cdot \hat q - i g(\sqrt{s})\, (\hat q^{\prime} \times
 \hat q) \cdot \vecsig \right) \hspace{0.7cm} (L=1)  \label{pwamp} \ ,
\end{eqnarray}
one defines the two amplitudes at tree level, $f_{-}^{\rm tree}$ ($L=1$, $J=1/2$) and 
$f_{+}^{\rm tree}$ ($L=1$, $J=3/2$), as
\begin{eqnarray}
    f_{+}^{\rm tree} &=& f+g \label{fg} \\
    f_{-}^{\rm tree} &=& f-2g \nonumber \ ,
\end{eqnarray}
with
\begin{eqnarray}
    f_{ij}(\sqrt{s}) &=& {1 \over 3} \left\{ - C_{ij} {1 \over 
4 f^2}\, a_i\,
    a_j \left({1 \over b_i} + {1 \over b_j} \right) 
    + { D^{\Lambda}_i D^{\Lambda}_j \left(1+{q_i^0 \over M_i} \right) 
    \left(1+{q_j^0 \over M_j} \right) \over \sqrt{s} - \tilde M_\Lambda} 
    \right.  \nonumber \\
    && \left.  + { D^{\Sigma}_i D^{\Sigma}_j \left(1+{q_i^0 \over M_i}
    \right) \left(1+{q_j^0 \over M_j} \right) \over \sqrt{s} - \tilde
    M_\Sigma} 
    + {2 \over 3} {D^{\Sigma^{*}}_i D^{\Sigma^{*}}_j \over
    \sqrt{s} - \tilde M_\Sigma^{*}} \right\} q_{i} q_{j}
    \label{f1}\\
    g_{ij}(\sqrt{s}) &=& {1 \over 3} \left\{  C_{ij} {1 \over 4 
f^2}\, a_i\,
    a_j \left({1 \over b_i} + {1 \over b_j} \right) 
    - { D^{\Lambda}_i D^{\Lambda}_j \left(1+{q_i^0 \over M_i} \right) 
    \left(1+{q_j^0 \over M_j} \right) \over \sqrt{s} - \tilde M_\Lambda} 
    \right.   \nonumber \\
    && \left.  - { D^{\Sigma}_i D^{\Sigma}_j \left(1+{q_i^0 \over M_i}
    \right) \left(1+{q_j^0 \over M_j} \right) \over \sqrt{s} - \tilde
    M_\Sigma} + {1 \over 3} {D^{\Sigma^{*}}_i D^{\Sigma^{*}}_j \over
    \sqrt{s} - \tilde M_\Sigma^{*}} \right\} q_{i} q_{j} \label{g1} \ ,
\end{eqnarray}
where $i,j$ are channel indices. 
Using Eq.~(\ref{BSeq}), one obtains
\begin{eqnarray}
    f_{+} &=& [1-f_{+}^{\rm tree} G ]^{-1} f_{+}^{\rm tree}  \ ,
    \label{fs} \\
    f_{-} &=& [1-f_{-}^{\rm tree} G ]^{-1} f_{-}^{\rm tree} \ .  \nonumber 
\end{eqnarray}
These two equations are analogous to solving the Bethe-Salpeter equation. The  $\Sigma^*$ pole  for $I=1$
is contained in the  $f_{+}$ amplitude while  the $f_{-}$ amplitude includes the $\Lambda$ and $\Sigma$ poles for
$I=0$ and $I=1$, respectively. As mentioned in Ref.~\cite{Jido:2002zk}, the 
unitarization procedure will shift the mass from a starting bare 
mass $\tilde M_{\Lambda}$, $\tilde M_{\Sigma}$, $\tilde M_{\Sigma^*}$  to the physically observed mass. 
  
As one can see from these equations, 
the amplitudes $f_{+}^{\rm tree}$, $f_{-}^{\rm  tree}$ in the 
diagonal meson-baryon channels contain the factor $\vecq\,^{2}$, with 
$\vecq$ the 
on-shell c.m.  momentum of the meson in this channel. For transition 
matrix elements from channel $i$ to $j$ the corresponding factor is 
$q_{i}q_{j}$, where the energy and momentum of the meson in a certain 
channel are given by the expressions 
\begin{equation}
    E_{i} = { s + m_{i}^{2} - M_{i}^{2} \over 2 \sqrt{s}} \ ; 
    \hspace{0.5cm} q_{i} = \sqrt{E_{i}^{2} - m_{i}^{2}} \ , 
\end{equation}
which also provide the analytical extrapolation below the threshold 
of the channel, where $q_i$ becomes purely imaginary. 

\section{In-medium \mbox{\boldmath{$\bar KN$}} interaction}
\label{medium}

The properties of the $\bar K$ in the nuclear medium are
obtained by incorporating the corresponding medium
modifications in the effective $\bar K N$ interaction. 

\subsection{Pauli blocking and self-energy effects}

One of the sources of density dependence comes from the Pauli principle, which prevents the
scattering to intermediate nucleon states below the Fermi momentum. This is
implemented by replacing the free nucleon propagator in the loop function by the
corresponding in-medium one.

Another
source of density dependence is related to the fact that all mesons and baryons
in the intermediate loops interact with the nucleons of the Fermi sea and their
properties are modified with respect to those in free space. 

The binding effects on the baryons are taken within a mean-field approach
consisting in adding, to the single-particle energies,  a momentum-independent
potential. In the case of nucleons, we take $U=U_0 \rho/\rho_0$, where
$\rho_0=0.17$ fm$^{-3}$ is normal nuclear matter density, and $U_0=-70$ MeV ,
which is in agreement with  numerous nuclear matter calculations with realistic
interactions \cite{Baldo,Brockmann:1990cn}. On the other hand, we use the
parametrization of Ref.~\cite{Millener:1988hp} for $\Lambda$,
$U^{\Lambda}=A\rho+B\rho^{\gamma}$ with $A=-340 \ {\rm fm^{-3}}$, $B=1087.5 \
{\rm fm^6}$ and $\gamma=2$. This parametrization shows a saturation behaviour
and can be used in the study of densities beyond $\rho_0$. Finally, since the
situation for the $\Sigma$-nucleus potential is unclear, we use for our
calculations the attractive potential $U^{\Sigma}=-30 \rho/\rho_0$ MeV, as
commonly accepted for low densities \cite{Batty:1978sb,Batty79}. One of the
outputs of the present study is the $\Lambda$ and $\Sigma$ self-energies in the
medium, which we obtain by looking at the shift of their respective poles in
the scattering matrix. A good degree of self-consistency is obtained between the
input $\Lambda$ and $\Sigma$ potentials and their corresponding output, 
which agree within 10 \% as will be seen in Sect.~\ref{results}.

The nuclear medium effects on the mesons will be included through the
corresponding self-energy. We will consider the dressing of the ${\bar K}$ and
$\pi$ mesons. The $\pi$ self-energy can be found in
Refs.~\cite{Ramos:1999ku,Ramos:1994xy}. It consists of a small 
$s$-wave part plus a $p$-wave part, which is constructed by allowing the pion
to couple to particle-hole, $\Delta$-hole and two-particle-hole excitations
modified by nuclear short-range correlations. 
The $\bar K$ self-energy is obtained from the $s$- and $p$-wave contributions
to the in-medium ${\bar K}N$
amplitude as explicitly shown in Sect.~\ref{sec:self}.

From the meson self-energies,
we can construct the dressed meson propagator ($i=\bar
K$,$\pi$)
\begin{equation}
D_i(q^0,\vec{q},\rho) = \frac{1}{(q^0)^2-{\vec q\,}^2 - m_i^2 -
\Pi_i(q^0,\vec{q},\rho)} \ ,
\end{equation}
and the corresponding spectral density
\begin{equation}
S_i(q^0,{\vec
q},\rho)= -\frac{1}{\pi} {\rm Im}\, D_i(q^0,{\vec q},\rho) =
-\frac{1}{\pi}\frac{{\rm Im} \Pi_i(q^0,\vec{q},\rho)}
{\mid (q^0)^2-\vec{q}\,^2-m_i^2-
\Pi_i(q^0,\vec{q},\rho) \mid^2} \ .
\label{eq:spec}
\end{equation}

\subsection{{\mbox{\boldmath \it s}}-wave in-medium amplitudes}

The calculation of the in-medium amplitudes requires a similar unitarization
procedure as that performed in free space. A key simplification in 
evaluating the free space
amplitudes was the factorization of the on-shell interaction kernel
out of the loop function. In this section we will show that the
on-shell factorization is still valid for the in-medium $s$-wave amplitudes.

\begin{figure}
   \begin{center}
   \epsfxsize=10cm
   \epsfbox{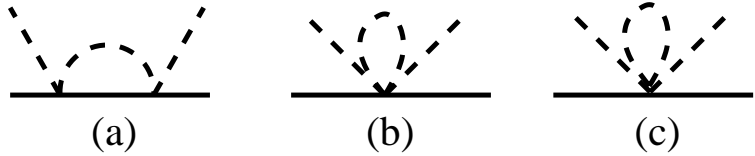}
   \end{center}
   \caption{ On-shell (a), off-shell (b) and tadpole (c) contributions for the $s$-wave amplitude in free space. \label{fig1}}  
\end{figure}

We first recall the arguments given in Ref.~\cite{Oset:1997it} to justify the validity
of the on-shell factorization in free space. Taking the structure of the
$s$-wave kernel given in Eq.~(\ref{swa2}), the off-shell dependence of the
loop function of Fig.~\ref{fig1}(a) from the two vertices goes as 
\begin{equation}
(k^0+q^0)^2=(2\,k^0+q^0-k^0)^2=(2\,k^0)^2+4\,k^0\,(q^0-k^0)+(q^0-k^0)^2 \ .
\label{eq:offshell}
\end{equation}
The first term on the right-hand side of this equation accounts for the on-shell
contribution in the vertices. The second and third terms cancel the
intermediate baryon propagator in the loop in the heavy baryon approach $(p^0 \approx
E(\vec{P}-\vec{q}\,))$, which becomes $(k^0-q^0)^{-1}$, as can be seen from
Eq.~(\ref{eq:gprop}). The resulting off-shell
contribution has the structure of Fig.~\ref{fig1}(b), which will be
conveniently canceled by a tadpole term, Fig.~\ref{fig1}(c), in a suitable
renormalization scheme.

\begin{figure}
   \begin{center}
   \epsfxsize=10cm
   \epsfbox{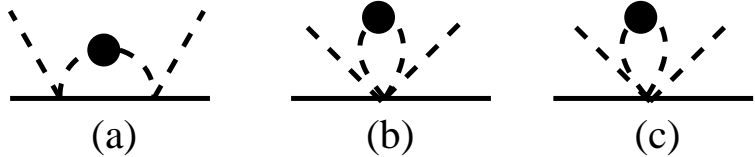}
   \end{center}
   \caption{  On-shell (a), off-shell (b) and tadpole (c) contributions for the $s$-wave amplitude including self-energy insertions.\label{fig2}}  
\end{figure}

In the nuclear medium, when we make self-energy attachments in the meson line,
we find contributions as that shown in Fig.~\ref{fig2} and, given the 
structure of the
off-shell part, Fig.~\ref{fig2}(b), and the tadpole, Fig.~\ref{fig2}(c), 
the cancellation that we had before still holds.
This justifies the use of the on-shell vertices in the medium for the $s$-wave.
Actually, the changes due to Pauli blocking in the nucleon line for the
off-shell terms are shown to vanish below. Indeed, the Pauli blocking correction
to the loop integral of diagram \ref{fig1}(a) from the first off-shell
contribution of Eq.~(\ref{eq:offshell}) is:
\begin{eqnarray}
\delta G_{l}^{\rm Pauli} &=& i \, \int \frac{d^4 q}{(2 \pi)^4} \,
\frac{M_l}{E_l
(\vec{P}-\vec{q}\,)} \,(q^0-k^0) \, \frac{1}{q^2 - m^2_l + i \epsilon}  \times
\nonumber \\
&& \left\{
\frac{1-n(\vec{P}-\vec{q}\,)}{P^0 - q^0 - E_l (\vec{P}-\vec{q}\,) +
i \epsilon} +
\frac{n(\vec{P}-\vec{q}\,)}{P^0 - q^0 - E_l (\vec{P}-\vec{q}\,) - i
\epsilon} \right.\nonumber \\
&& \left. -\frac{1}{P^0 - q^0 - E_l (\vec{P}-\vec{q}\,) + i
\epsilon} \right\} \ ,
\label{eq:gpauli0}
\end{eqnarray}
with $P^0=k^0+p^0$. Since the Pauli blocking corrections are only operative at the nucleon pole, one finds
\begin{eqnarray}
\delta G_{l}^{\rm Pauli} &=& i \, \int \frac{d^4 q}{(2 \pi)^4} \,\frac{M_l}{E_l
(\vec{P}-\vec{q}\,)} \,  (q^0-k^0) \, \frac{1}{q^2 - m^2_l + i \epsilon}   \times \nonumber \\
&& 2 \, i \, \pi \, n(\vec{P}-\vec{q}\,)
\,\delta(p^0+k^0-q^0-E_l(\vec{P}-\vec{q}\,)) \ .
\label{eq:gpauli1}
\end{eqnarray}
Since $p^0-E_l(\vec{P}-\vec{q}\,) \approx 0$ in the heavy-baryon approach, the
delta function forces the factor $(q^0-k^0)$ to be zero and the correction
$\delta G_l^{\rm Pauli}$ vanishes. An identical argument holds
for the other off-shell term, which is proportional to $(q^0-k^0)^2$.  

Generalizing these findings to all orders, it is concluded that the on-shell
factorization can be applied for the in-medium $s$-wave amplitudes and the loop
function is simply the integral of a meson and a baryon propagator, conveniently
modified by self-energy insertions and binding corrections, namely:
\begin{eqnarray}
&&G_{\bar K N}^s(P^0,\vec{P},\rho)=
\int_{\mid
{\vec q}\, \mid < q_{\rm max}^{\rm lab}} \frac{d^3 q}{(2 \pi)^3}
\frac{M_N}{E_N (\vec{P}-\vec{q}\,)} \times  \nonumber \\
&&\left[
\int_0^\infty d\omega
 S_{\bar K}(\omega,{\vec q},\rho)  
\frac{1-n(\vec{P}-\vec{q}\,)}{P^0- \omega
- E_N (\vec{P}-\vec{q}\,)
+ i \epsilon} \right. \nonumber \\
&&+ \left. \int_0^\infty d\omega
 S_{K}(\omega,{\vec q},\rho) 
\frac{n(\vec{P}-\vec{q}\,)}
{P^0+ \omega - E_N(\vec{P}-\vec{q}\,)+i \epsilon } \right] \ ,
\label{eq:gmedkaon}
\end{eqnarray}
for ${\bar K}N$ states and
\begin{eqnarray}
G_{\pi Y}^s(P^0,\vec{P},\rho)&= &
\int_{\mid
{\vec
q}\, \mid < q_{\rm max}^{\rm lab}} \frac{d^3 q}{(2 \pi)^3}
\frac{M_Y}{E_Y (\vec{P}-\vec{q}\,)} 
\int_0^\infty d\omega
 S_\pi(\omega,{\vec q},\rho) \nonumber
\\
& \times & 
\frac{1}{P^0- \omega
- E_Y (\vec{P}-\vec{q}\,)
+ i \epsilon}  \ ,
\label{eq:gmedpion}
\end{eqnarray}
for $\pi \Lambda$ or $\pi \Sigma$ states, where $P=(P^0,\vec{P}\,)$ is the total
four-momentum and ${\vec q}$ is the meson momentum in the laboratory system.

For $\eta \Lambda$, $\eta \Sigma$ and $K \Xi$ states, 
no self-energy insertions are incorporated in the meson lines and
we can use the loop integral in free space of
Eq.~(\ref{eq:gprop}), modified by 
including the binding potential in the baryon energy.

The in-medium $s$-wave amplitudes are then obtained by solving the coupled-channel
Eq.~(\ref{BSeq})
with these medium modified loop functions. 

\subsection{{\mbox{\boldmath \it p}}-wave in-medium amplitudes}

\begin{figure}
   \begin{center}
   \epsfxsize=10cm
   \epsfbox{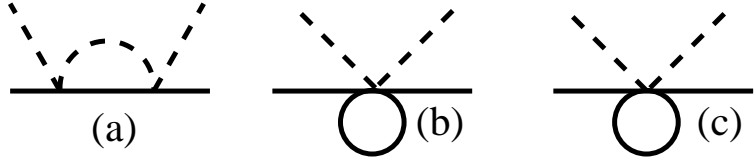}
   \end{center}
   \caption{On-shell (a), off-shell (b) and tadpole (c) contributions for 
 the  $p$-wave amplitude in free space. \label{fig3}}  
\end{figure}

The situation is different for the $p$-wave amplitudes. The $\vec{q}\,^2$
dependence from the vertices in the one-loop contribution to the $p$-wave
amplitude can be separated
into an on-shell part, $\vec{q}\,^2_{\rm on}$, and an off-shell part,
$\vec{q}\,^2-\vec{q}\,^2_{\rm on}$. In free space, the one-loop
contribution, splitted into its on-shell and off-shell parts, is represented by the
diagrams of Fig.~\ref{fig3}(a) and (b), respectively. 
The factor $\vec{q}\,^2-\vec{q}\,^2_{\rm on}$ can
be shown to cancel the meson propagator (see Ref.~\cite{Cabrera:2002hc})
and, hence, the off-shell contribution, Fig.~\ref{fig3}(b), 
is canceled by the tadpole term of Fig.~\ref{fig3}(c).

\begin{figure}
   \begin{center}
   \epsfxsize=10cm
   \epsfbox{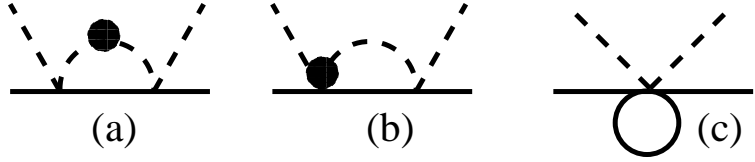}
   \end{center}
   \caption{  On-shell (a), off-shell (b) and tadpole (c) contributions for 
   the $p$-wave amplitude including self-energy insertions. \label{fig4}}  
\end{figure}

In the medium, when the meson propagator is dressed, one encounters the
situation of Fig.~\ref{fig4}, where in the diagram \ref{fig4}(b) it is seen that
the factor $\vec{q}\,^2-\vec{q}\,^2_{\rm on}$  only cancels one of the 
two intermediate meson propagators and, furthermore, there are no 
medium corrections for the
self-energy insertion in the inexistent intermediate mesons of the tadpole term
in Fig.~\ref{fig4}(c). Hence, in the medium, we do not find the cancellation
between the off-shell part and the tadpole term that applied to the
$s$-wave amplitudes. However, there is no problem if we add, to the free
loop function (for which the on-shell prescription is valid) 
the medium corrections calculated
using the full off-shell $\vec{q}\,^2$ contribution of the vertices, according
to the following replacement
\begin{eqnarray}
\label{gtogmedio}
G_l^p(s) \to G_l^p(s) + \frac{1}{\vec{q}\,^2_{on}} \lbrack I_{\rm med}(s) -
I_{\rm free}(s)
\rbrack \ ,
\end{eqnarray}
where
\begin{eqnarray}
\label{Ifunctions}
I_{\rm med}(s) &=& i \int \frac{d^{4}q}{(2\pi)^{4}} \vec{q}\,^2 D_M(q) G_B(P-q)
\nonumber\\
I_{\rm free}(s) &=& i \int \frac{d^{4}q}{(2\pi)^{4}} \vec{q}\,^2 D^0_{M}(q) 
G^0_{B}(P-q) \ ,
\end{eqnarray}
where $G_l^p$ is the free loop of Eq.~(\ref{eq:gprop}), $D_M(q)$ and $G_B(P-q)$
stand for the meson and baryon propagator in the medium, respectively,
while $D^0_M(q)$ and $G^0_B(P-q)$ correspond to those in free space.  
The tadpole terms have been assumed to cancel in the difference on the r.h.s. of
Eq.~(\ref{gtogmedio}).

\begin{figure}
   \begin{center}
   \epsfxsize=4cm
   \epsfbox{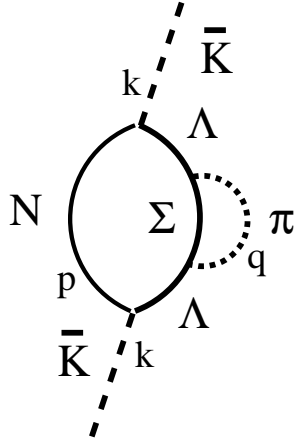}
   \end{center}
   \caption{ $p$-wave contribution to the ${\bar K}$ self-energy  \label{kaondiagram}}
\end{figure}

In order to see explicitly the problems of the on-shell factorization, let us
evaluate the contribution to the antikaon self-energy 
shown in Fig.~\ref{kaondiagram}, where the ${\bar K}$ couples to a $\Lambda
N^{-1}$ excitation and the $\Lambda$ line contains a $\Sigma \pi$ self-energy
insertion. The corresponding expression for the self-energy is
\begin{eqnarray}
-i \, \Pi(k) &=& g_{N\Lambda K}^2 \, \int \frac{d^4p}{(2\pi)^4} (-2 \vec{k}\,^2)
\frac{i n(\vec{p}\,)}{p^0-E(\vec{p}\,)+i\varepsilon}\left( \frac{i}{p^0+k^0-
E_\Lambda(\vec{p}+\vec{k}\,) + i\varepsilon}\right)^2 \nonumber \\  
&& \times g_{\Lambda \Sigma \pi}^2  \int \frac{d^4q}{(2 \pi)^4} (-\vec{q}\,^2)
\frac{i}{p^0+k^0-q^0-E_\Sigma(\vec{p}+\vec{k}-\vec{q}\,)} 
\frac{i}{q^{0\,2}-\vec{q}\,^2-m_\pi^2+i\varepsilon}  \ .
\end{eqnarray}
Applying Cutkosky rules 
\begin{equation}
\begin{array}{lcl}
\Pi(k) &\rightarrow& 2 \, i \, {\rm Im} \Pi(k)  \\
G_N(p) &\rightarrow& 2\, i \, \theta(p^0)\, {\rm Im} G_N(p) \\
G_\Sigma(p+k-q) &\rightarrow& 2\, i \, \theta(p^0+k^0-q^0) {\rm Im}
G_\Sigma(p+k-q) \\
D_\pi(q) &\rightarrow& 2 \, i \, \theta(q^0) \,  {\rm Im} D_\pi(q)
\end{array}
\end{equation}
to evaluate the imaginary part corresponding to 
a cut producing $\Sigma \pi N^{-1}$ states,
we obtain
\begin{eqnarray}
{\rm Im} \Pi(k) &=& - g_{N\Lambda K}^2 g_{\Lambda \Sigma \pi}^2 \vec{k}\,^2 
\int \frac{d^3p}{(2\pi)^3} n(\vec{p}\,) 
\left( \frac{1}{p^0+k^0-
E_\Lambda(\vec{p}+\vec{k}\,)}\right)^2 \nonumber \\
&& \times \int \frac{d^3q}{(2\pi)^3} \frac{\vec{q}\,^2}{2\omega(\vec{q}\,)}
\delta(k^0+E_N(p)-\omega(\vec{q}\,) - E_\Sigma(\vec{p}+\vec{k}-\vec{q}\,)) \ .
\end{eqnarray}
If one blindly applies the on-shell factorization to this $p$-wave
contribution, the factor $\vec{q}\,^2$ would be replaced by
$\vec{q}\,^2_{\rm on}$ and taken out
of the integral over the running variable $\vec{q}$. For an 
incident antikaon energy, $k^0$, such that $m_\pi +
M_\Sigma < k^0 + M_N $ we would find $\vec{q}\,^2_{\rm on}> 0$. If
this value of the kaon energy is below the kaon mass, 
namely $k^0 + M_N < m_K + M_N$, then the
corresponding on-shell antikaon momentum would
fulfill $\vec{k}\,^2 < 0$ and, in this
case, we would end up having a positive contribution to the imaginary part of
the ${\bar K}$
self-energy, hence violating the principle of causality.
However, this would not be a real problem since, in the medium, one needs 
to keep the proper independence of the kaon energy-momentum variables, and this
amplitude will be corrected, as explained at the end of this section, by a
factor $k_{\rm lab}^2/{k^2}$, where $\vec{k}_{\rm lab}$ is the antikaon momentum in
the laboratory system, which is a well defined real quantity.
Having this correction in mind, the pathology would then occur
when $ k^0 + M_N  < m_\pi + M_\Sigma$, since in this case $\vec{q}\,^2_{\rm on} < 0$,
hence leaving a positive contribution to the imaginary part of the $\bar K$
self-energy for a kaon with momentum $\vec{k}_{\rm lab}$ and energy
$k^0 < m_\pi + M_\Sigma - M_N$.

\begin{figure}[htb]
\begin{minipage}{120mm}
\includegraphics[width=6cm]{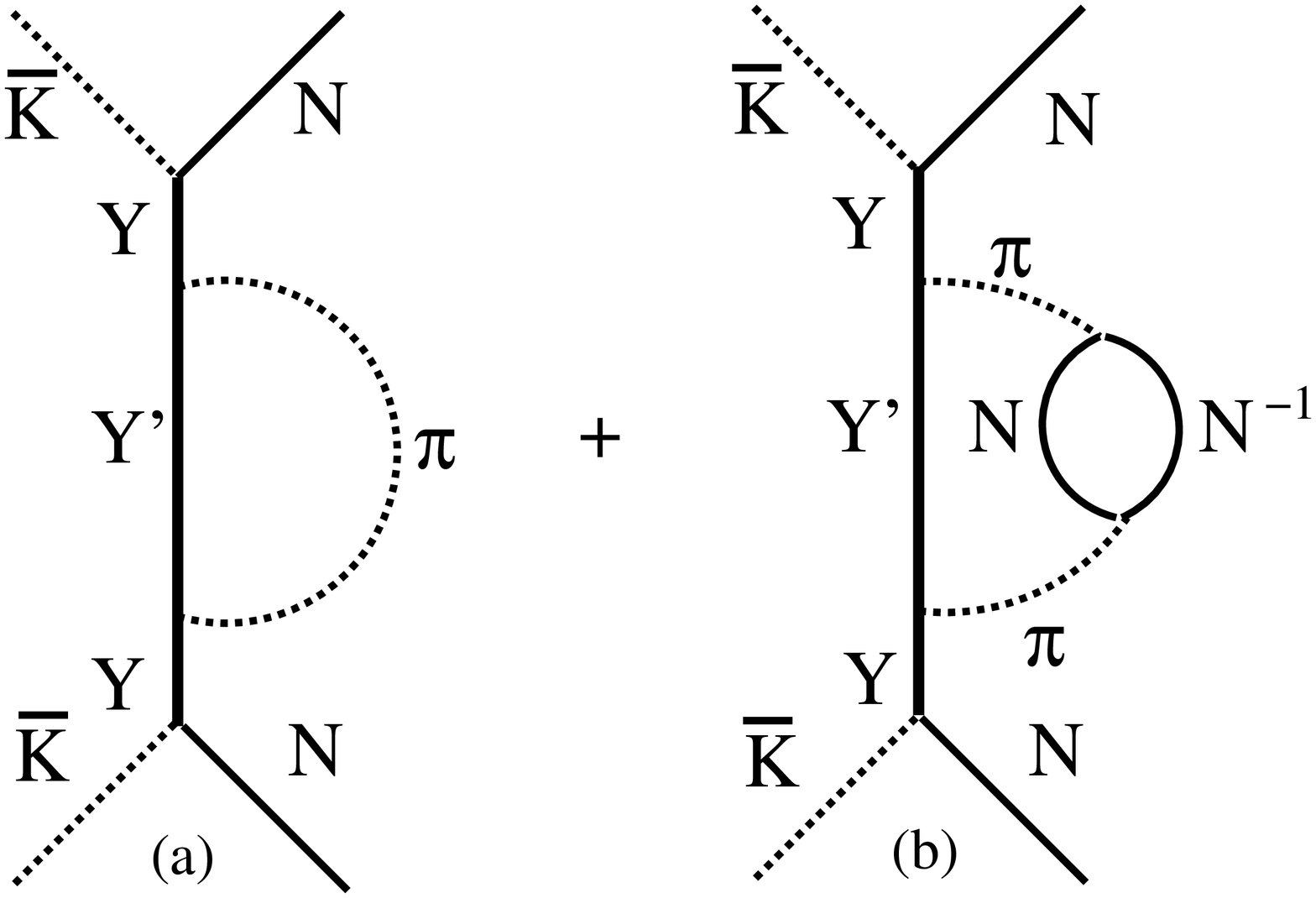}~~~~
\includegraphics[width=6cm]{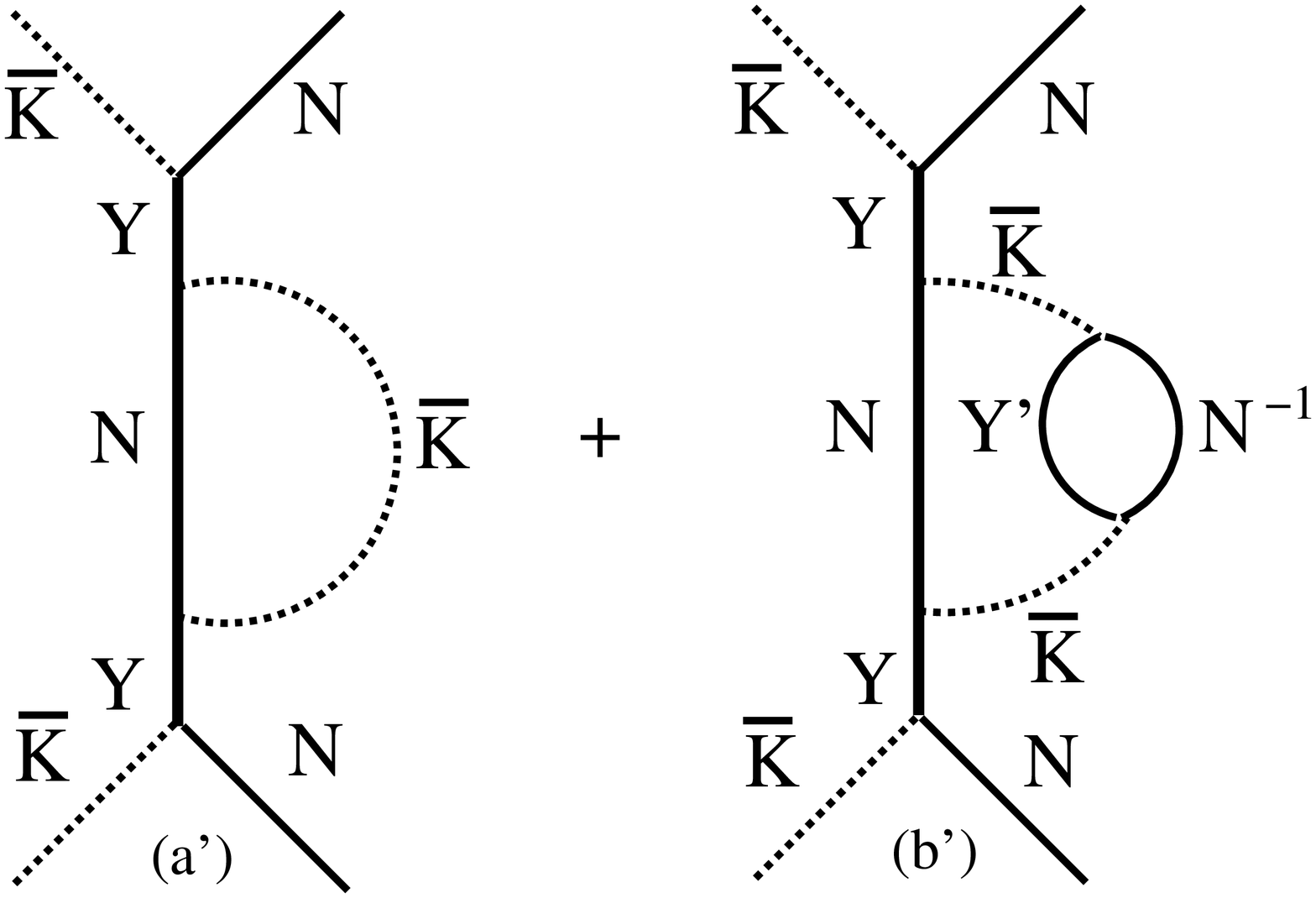}
 
   \caption{Diagrams contributing to the $\bar K N$ $p$-wave amplitude including intermediate pions (a,b) and intermediate kaons (a',b')\label{src}}  

\end{minipage}
\end{figure}

Another ingredient that has to be considered when dealing with $p$-wave
amplitudes in the medium is the effect of nuclear short-range correlations. 
Diagrams such as that of Fig.~\ref{src}
are present in the in-medium $p$-wave $\bar K N$ amplitude and, therefore, the
$\pi$ ($\bar K$)  propagators connecting a bubble to a baryonic line 
(see Figs.~\ref{src}(b) and (b')) or to another bubble should be modified in order to 
account for the fact that the nucleon-nucleon (hyperon-nucleon) interaction is not only
driven  by one-pion (one-kaon) exchange.

Let's take the case of pions explicitly. The contribution of the diagram of Fig.~\ref{src}(b)
reads
\begin{eqnarray}
&&F^4(\vec{q}\,) \vec{q} D^0(q^0,\vec{q}\,)
 \left( \frac{f_{\pi N N}}{m_{\pi}} \right)^2 U(q^0,\vec{q}\,) \vec{q}\,^2 
  D^0(q^0,\vec{q}\,) \vec{q} \nonumber \\
&&= F^4(\vec{q}\,)  \delta_{il} 
q_i  D^0(q^0,\vec{q}\,) \left( \frac{f_{\pi N N}}{m_{\pi}} \right)^2 
 U(q^0,\vec{q}\,) q_j  q_j  D^0(q^0,\vec{q}\,)  q_{l} ,
\label{src0}
\end{eqnarray}
where $U(q^0,\vec{q}\,)$ corresponds to the Lindhard function and
$F(\vec{q}\,)=\Lambda^2/(\Lambda^2+\vec{q}\,^2)$ with $\Lambda=1$~GeV.
The effect of
short-range correlations is incorporated by replacing the exchanged pion 
in Fig.~\ref{src}(b) by a fully correlated interaction, which is achieved by
the following replacement in Eq.~(\ref{src0})
\begin{equation}
F^2(\vec{q}\,)^2  q_i  q_j  D^0(q^0,\vec{q}\,) \rightarrow V_l \hat{q}_i 
 \hat{q}_j  +  V_t  (\delta_{ij}-\hat{q}_i \hat{q}_j) \ ,
\end{equation}
where
\begin{eqnarray} 
V_l&=&(\vec{q}\,^2 D^0(q^0,\vec{q}\, ) + g') F^2(\vec{q}\,),
 \hspace{0.5cm}  V_t = g'  F^2(\vec{q}\,) \nonumber \ ,
\end{eqnarray}
with $g'\simeq 0.6$ being the usual Landau-Migdal parameter \cite{Migdal:1978az}. 
Then Eq.~(\ref{src0}) reads
\begin{eqnarray}
&&\delta_{il}  \left[ V_l \hat{q}_i
 \hat{q}_j +  V_t (\delta_{ij} - \hat{q}_i \hat{q}_j) \right]
  \left( \frac{f_{\pi N N}}{m_{\pi}} \right)^2  U(q^0,\vec{q}\,)
    \left[ V_l \hat{q}_j \hat{q}_l +  V_t (\delta_{jl}  -  
    \hat{q}_j \hat{q}_l) \right] \nonumber \\
&& =\delta_{il} 
\left[ V_l^2 \hat{q}_i \hat{q}_l + V_t^2 (\delta_{il} - \hat{q}_i \hat{q}_l) 
\right]\left( \frac{f_{\pi N N}}{m_{\pi}} \right)^2  U(q^0,\vec{q}\,) \ .
\end{eqnarray}
Since the longitudinal and transverse components are decoupled from each other,
successive iterations lead to
\begin{eqnarray}
\frac{V_l^2 
\left( \frac{f_{\pi N N}}{m_{\pi}} \right)^2 U(q^0,\vec{q}\,)}
{1-V_l \left( \frac{f_{\pi N N}}{m_{\pi}} \right)^2 U(q^0,\vec{q}\,)}
 + 2 \frac{V_t^2 \left( \frac{f_{\pi N N}}{m_{\pi}} \right)^2 
  U(q^0,\vec{q}\,)}{ 1-V_t \left( \frac{f_{\pi N N}}{m_{\pi}} \right)^2 
    U(q^0,\vec{q}\,)}  \ .
\label{src1}
\end{eqnarray}
Using the fact that the Lindhard function and $p$-wave self-energy are related 
by $(f_{\pi N N}/m_{\pi})^2\,U(q^0,\vec{q}\,)=\Pi^{p}(q^0,\vec{q}\,)/ (\vec{q}\,^2
F^2(\vec{q}\,)) $ and summing the contribution of the first diagram,
Fig.~\ref{src}(a), the full dressed pion propagator without correlations
$D(q^0,\vec{q}\,)$ has to be substituted by
\begin{eqnarray}
&&F^2(\vec{q}\,) \vec{q}\,^2 D(q^0,\vec{q}\,) \rightarrow  
F^2(\vec{q}\,)
\vec{q}\,^2 \frac{1}{q^{0\,2}- \vec{q}\,^2-m_{\pi}^2-\Pi_{\pi}^{s}} + 
\nonumber \\
&& \frac{V_l^2 
 \Pi^{p}(q^0,\vec{q}\,)/(\vec{q}\,^2 F^2(\vec{q}\,))}{1-V_l
  \Pi^{p}(q^0,\vec{q}\,)/(\vec{q}\,^2 F^2(\vec{q}\,))}  
+ 2 \frac{V_t^2 \Pi^{p}(q^0,\vec{q}\,)/(\vec{q}\,^2 F^2(\vec{q}\,))}
{1-V_t \Pi^{p}(q^0,\vec{q}\,)/(\vec{q}\,^2 F^2(\vec{q}\,))} \ ,
\end{eqnarray} 
where the $s$-wave piece of the self-energy has been included in the definition
of the free pion
propagator. Alternatively, 
\begin{eqnarray}
&&F^2(\vec{q}\,) \vec{q}\,^2 D(q^0,\vec{q}\,) \rightarrow \nonumber \\
&& \frac{1-g' F^2(\vec{q}\,) {V_l}^{-1} + 
  g' \Pi^{p}(q^0,\vec{q}\,)/ \vec{q}\,^2}{{V_l}^{-1} - 
   \Pi^{p}/ (\vec{q}\,^2 F^2(\vec{q}\,))}  +  2 \frac{ V_t^2 
   \Pi^{p}(q^0,\vec{q}\,)/ (\vec{q}\,^2 F^2(\vec{q}\,)}
   {1- V_t \Pi^{p}(q^0,\vec{q}\,)/ (\vec{q}\,^2 F^2(\vec{q}\,))} \ .
\end{eqnarray}
A similar expression is obtained for the antikaon propagator when
the diagram of Fig.~\ref{src}(b') and its subsequent
iterations, all corrected by the short-range effects of the hyperon-nucleon
interaction, are added to the diagram of
Fig.~\ref{src}(a').
 
The in-medium $p$-wave amplitudes are finally obtained by solving the
same equations as in free space, Eqs.~(\ref{fs}), taking the tree amplitudes from
Eqs.~(\ref{fg})-(\ref{g1}), evaluated with on-shell momentum values depending on
$\sqrt{s}$, 
and using the in-medium
meson-baryon propagator of Eq.~(\ref{gtogmedio}), which incorporates the proper
$\vec{q}\,^2$ dependence in the medium corrections to the $p$-wave amplitudes,
as well as Pauli blocking effects, dressing of the mesons, binding potentials
for the baryons and  nuclear short-range correlations. 

Finally, the amplitudes
need to be corrected to incorporate, in the external states, 
the proper off-shell momentum, which we write for convenience in
terms of the laboratory variables. Furthermore, in going from c.m. to lab
momenta, the vertex recoil factors also need to be corrected.
Technically this is implemented by the following expression:
\begin{eqnarray} (f_{+})_{ij}^{\rm med} &=& ([1-f_{+}^{\rm tree} G ]^{-1}
f_{+}^{\rm tree})_{ij} \frac{(q_{\rm lab})_i  (q_{\rm lab})_j}
{(q_{\rm on})_i  (q_{\rm on})_j} \left(
1-\frac{(q^0_{\rm lab})_i}{M_{\Sigma^*}} \right) \left(
1-\frac{(q^0_{\rm lab)_j}}{M_{\Sigma^*}} \right) \nonumber \\ 
(f_{-})_{ij}^{\rm med} &=&
([1-f_{-}^{\rm tree} G ]^{-1} f_{-}^{\rm tree})_{ij}  \frac{(q_{\rm lab})_i 
(q_{\rm lab})_j}{(q_{\rm on})_i  (q_{\rm on})_j} 
\frac{\left( 1-\frac{(q^0_{\rm lab})_i}{2\overline{M}}
\right) \left( 1-\frac{(q^0_{\rm lab})_j}{2 \overline{M}}\right)} {\left(
1+\frac{(q^0_{\rm on})_i}{2 M} \right) \left( 1+\frac{(q^0_{\rm on})_j}{2 M} \right)} \ ,
\end{eqnarray} 
where $q_{\rm on}$ and $q_{\rm lab}$ stand, respectively, for the on-shell
momentum and the momentum in the laboratory system, $M$ is the nucleon
mass and $\overline{M}$ is the average $\Sigma$-$\Lambda$ mass.
Actually, for the $f_{-}$ amplitude we could separate between the $I=0$ and $I=1$
amplitudes and introduce the recoil corrections with the corresponding $\Lambda$
and $\Sigma$ masses, respectively. However, given
the similar masses of the $\Sigma$ and $\Lambda$, the use of an averaged mass
induces errors of less than 1\%.
We note that, by correcting the external
meson-baryon channels with the relativistic recoil factors appropriate for
pole-type terms, we are also performing an unnecessary correction on 
the contact term of the $p$-wave amplitudes, the 
first term in Eqs.~(\ref{f1}),(\ref{g1}). However, this induced error 
is negligible since this term is very small compared to the $\Lambda$, $\Sigma$ and $\Sigma^*$
pole contributions to the $p$-wave amplitude. 

\subsection{\mbox{\boldmath{$\bar K$}} self-energy}
\label{sec:self}
The $\bar K$ self-energy in  dense nuclear matter is obtained by summing the
in-medium $\bar K$ interaction $T_{\bar K N}$ for $s$- and $p$- waves over the
Fermi sea of nucleons according to
\begin{equation}
\Pi_{\bar{K}}(q^0,{\vec q},\rho)=4\int \frac{d^3p}{(2\pi)^3}\,
n(\vec{p}\,) \,  T_{\bar K N}(P^0,\vec{P},\rho) \ ,
\label{eq:selfka}
\end{equation}
where $P^0=q^0+E(p)$ and $\vec{P}=\vec{q}+\vec{p}$ are the total energy and
momentum in the lab frame and the values ($q^0$,$\vec{q}\,$) stand  for the
energy and momentum of the $\bar K$ in this frame. Note that the
$\bar K$ self-energy must be determined self-consistently, since it is 
obtained from the in-medium
interaction which uses $\bar K$ propagators which themselves include the
self-energy being calculated.

We finally remark that the recoil corrections discussed at the end of the
previous section assumed the nucleons to be at rest in the
laboratory frame. However, in the evaluation of the ${\bar K}$ self-energy, 
one works in a frame where the nuclear Fermi sea is at rest and the recoil
corrections will then have to take into account the Fermi
motion of nucleons. These corrections, which 
induce a small contribution to the
$s$-wave amplitudes,
are studied in detailed in
Ref.~\cite{Garcia-Recio:1988bs} and are incorporated in our calculations. For
completeness, we reproduce here the expression for these $p$-wave induced
Fermi motion corrections to the $s$-wave self-energy:
\begin{eqnarray}
\Pi^{(s,{\rm ind})}_{\bar K}(q^0,\vec{q},\rho) &=&
\frac{3}{5} k_F^2 (q^{0})^2 \left[
\frac{1}{4} \left(\frac{1}{M_N} + \frac{1}{M_\Lambda}\right)^2
D^2_{{\bar K} N \Lambda} U_\Lambda(q^0,\vec{q}\,) \right. \nonumber \\
& + & \left. \frac{1}{4} \left(\frac{1}{M_N} + \frac{1}{M_\Sigma}\right)^2
D^2_{{\bar K} N \Sigma} U_\Sigma(q^0,\vec{q}\,) 
+  \left(\frac{1}{M_{\Sigma^*}}\right)^2
D^2_{{\bar K} N \Sigma^* } U_{\Sigma^*}(q^0,\vec{q}\,) \right] \ 
\end{eqnarray}
where $U_Y$ stands for the hyperon-hole Lindhard function and 
$D_{{\bar K} N Y}$
is the coupling constant of the ${\bar K}NY$ vertex.

\section{Results}
\label{results}

\begin{figure}
\begin{center}
    \includegraphics[width = \textwidth]{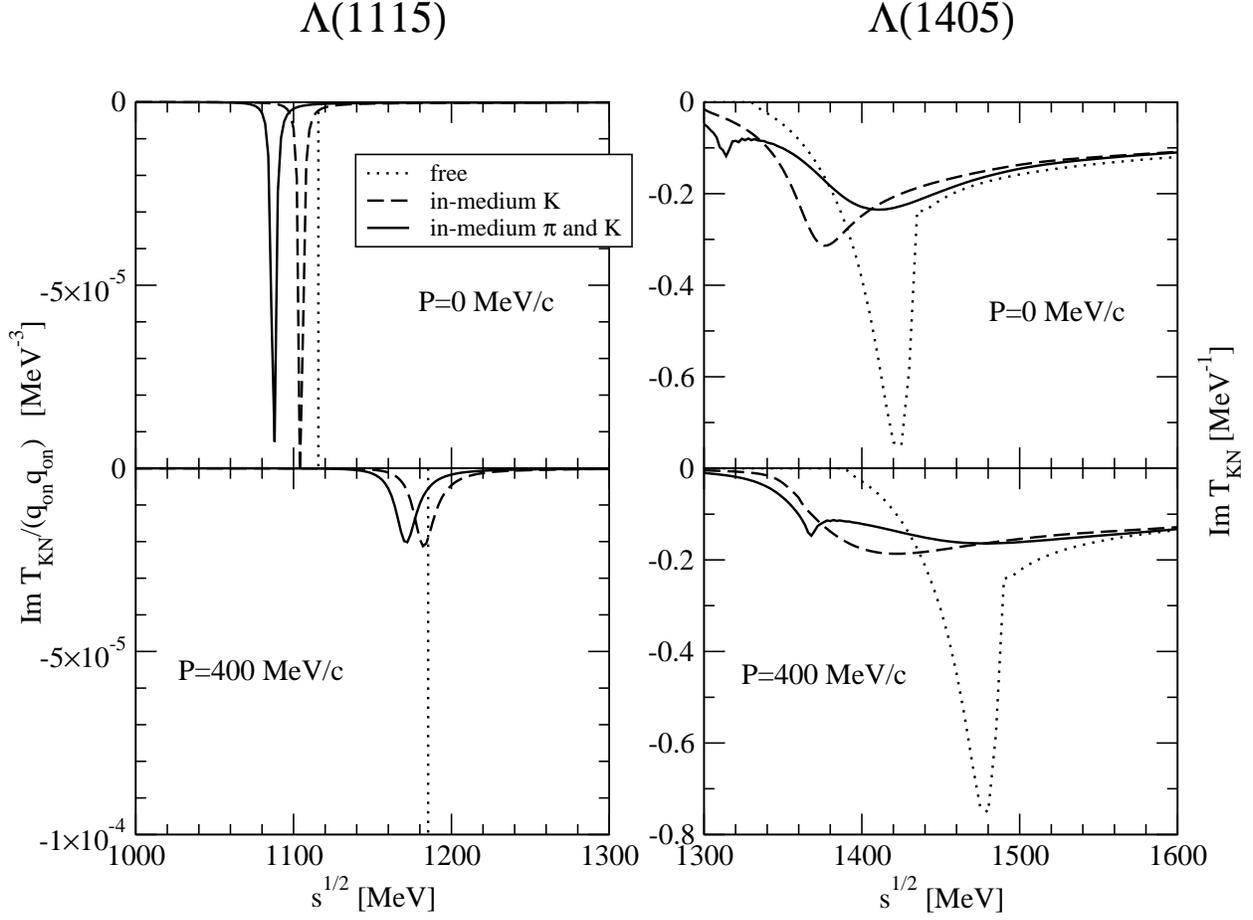}
\caption{Imaginary part of the $I$$=$$0$, $J^P$$=$$1/2^+$, ${\bar K}N \to {\bar K}N$ amplitude as a function
of $\sqrt{s}$ for two values of the momentum $\mid \vec{P}\,\mid$, in the $s$- (right panel) and $p$- (left panel) waves.
The free space
amplitudes (dotted lines) are compared with the in-medium amplitude at
$\rho$$=$$\rho_0$ dressing only antikaons (dashed lines) or dressing also the pions
(solid lines).}
\label{fig:lambda}
\end{center}
\end{figure}

In Fig.~\ref{fig:lambda} we present our results for the $I=0$
$J^P$$=$$1/2^+$ resonances, the $s$-wave $\Lambda(1405)$ (right panel)
and the $p$-wave $\Lambda(1115)$ (left panel). We show the imaginary
part of the ${\bar K}N \to {\bar K}N$ amplitude as a function of
$\sqrt{s}$ for two values of the momentum $P$. The $p$-wave amplitude is divided
by the square of the ${\bar K}N$ on-shell momentum corresponding 
to each value of
$\sqrt{s}$ .
The free space amplitudes (dotted lines) are compared with the
in-medium ones at $\rho=\rho_0=0.17$ fm$^{-3}$ for two
approximations: dressing the antikaons self-consistently (dashed
lines) and considering also the in-medium effects on the
properties of the pions (solid lines).  The apparent width of
the $P=0$ $\Lambda(1115)$ is fictitious and comes from a small
width inserted in the $\Lambda$ and $\Sigma$ pole driving terms to
facilitate the convergence of the self-consistent calculations. At
finite total momentum the width is physical since the $\Lambda$
can excite, via its interactions with nucleons in the Fermi sea, intermediate
$\Lambda N N^{-1}$ states having the same total momentum $P$.
We find, similarly to what is found in the model of
Ref.~\cite{lutz-korpa02}, that the $\Lambda(1115)$ acquires an
attractive shift of about 10 MeV when the self-consistent dressing
of the antikaons is considered. If the in-medium pion self-energy
is also incorporated, the $\Lambda(1115)$ develops an attraction
three times larger, of 28 MeV, in accordance to what is demanded
by hypernuclear spectroscopic data \cite{hyper} and also to what
is obtained from nuclear matter microscopic calculations using the
recent meson-exchange $YN$ potentials \cite{julich,nijmegen,Vidana:2001rm,haidenbauer}. The reason lies in 
that the dressing of the pions implicitly incorporates,
through the coupling of the pion to $ph$ states, an important piece
of the $\Lambda N$ interaction, namely the $\Lambda N \to \Sigma
N$ transition potential mediated by pion-exchange.

The $\Lambda(1405)$ shows the features that were already observed
in our earlier paper \cite{Ramos:1999ku} and which we summarize
here. Since the self-consistent dressing lowers the antikaon mass
by about 50 MeV and, in addition, a stronger binding
potential is taken for nucleons than for hyperons, the $\bar{K}N$
threshold gets effectively lowered in the medium and the
$\Lambda(1405)$ is dynamically generated at a lower value of
$\sqrt{s}$ (dashed line). When pions are dressed new
channels are available, such as $\Lambda N N^{-1}$ or $\Sigma N
N^{-1}$, so the $\Lambda(1405)$ gets strongly diluted and the peak
appears very close to the free space position. We note the cusp
effect that appears at an energy corresponding to the in-medium
$\pi \Sigma$ threshold as a consequence of the opening of this
channel on top of the already opened $\Lambda N N^{-1}$ and
$\Sigma N N^{-1}$ ones.

\begin{figure}
\begin{center}
    \includegraphics[width = \textwidth]{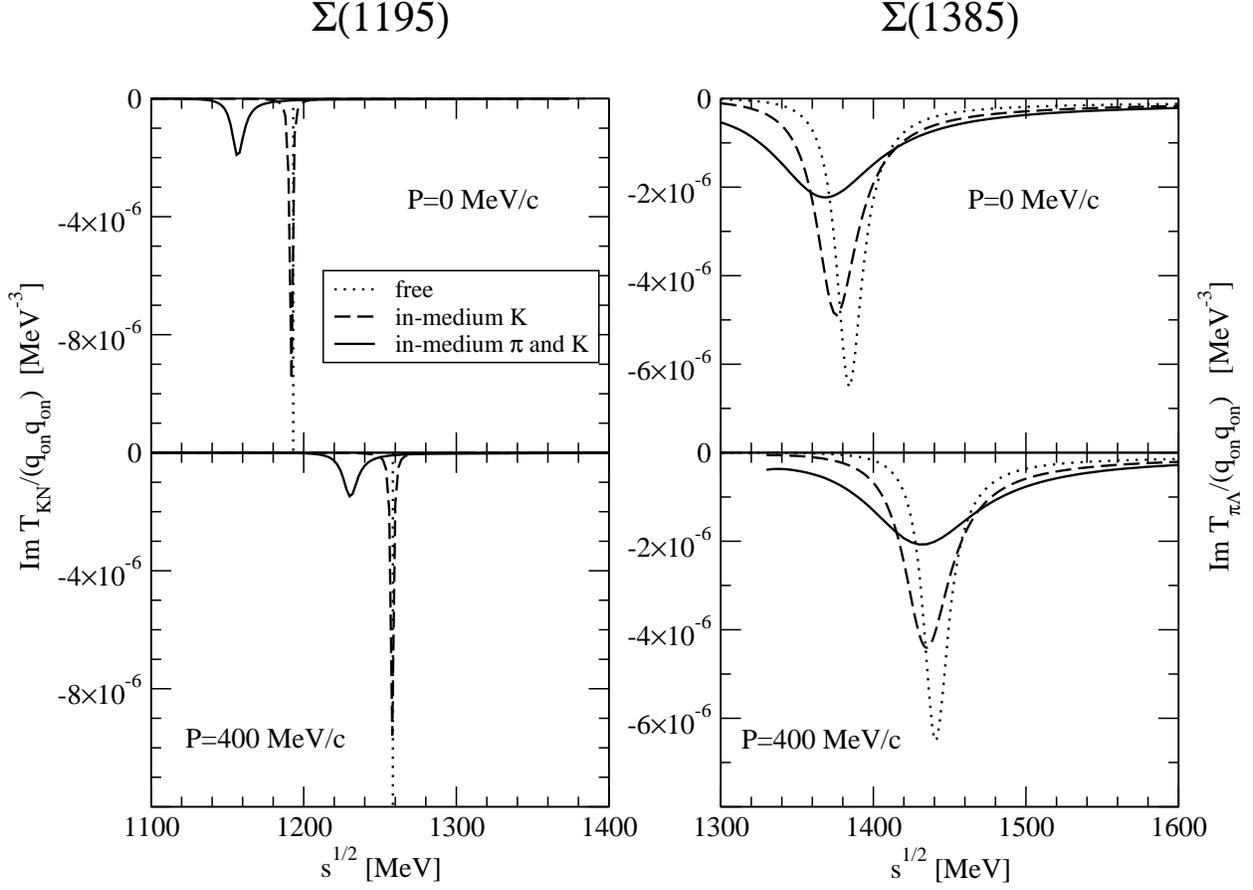}
\caption{Imaginary part of the $I$$=$$1$, $J^P$$=$$1/2^+$, $L$$=$$1$
${\bar K}N \to {\bar K}N$ amplitude and of the $I$$=$$1$, $J^P$$=$$3/2^+$, $L$$=$$1$
$\pi\Lambda \to \pi\Lambda$ amplitude  as functions
of $\sqrt{s}$ for two values of the momentum $\mid \vec{P}\,\mid$.
The free space
amplitudes (dotted lines) are compared with the in-medium amplitudes at
$\rho$$=$$\rho_0$ dressing only antikaons (dashed lines) or dressing also the pions
(solid lines).}
\label{fig:sigma}
\end{center}
\end{figure}

The $I=1$ resonances are shown in Fig.~\ref{fig:sigma}. The left
panels display the imaginary part of the ${\bar K}N \to {\bar K}N$
amplitude divided by the square of the ${\bar K}N$ 
on-shell momentum in the $J=1/2^+$  $p$-wave corresponding to the
$\Sigma(1195)$, while the right panels show the imaginary part of
the $\pi\Lambda \to \pi\Lambda$ amplitude divided by
the square of the $\pi\Lambda$ on-shell momentum in the $J=3/2^+$ $p$-wave
corresponding to the $\Sigma(1385)$. The width of the
$\Sigma(1195)$ is a reflection of the decay channel $\Sigma N \to
\Lambda N$, which is incorporated as soon as the antikaon or/and
the pions are dressed. However, we expect pion dressing to play a
stronger role due to the larger value of the coupling
constants involved and to the relative size between the pion and
kaon propagators. We indeed observe that pion dressing produces
an additional attraction of about 35 MeV to that obtained with the
approximation of dressing only the antikaons, which only produces
an attraction of 2 MeV. This last value is in contrast with the 10
MeV attraction found in Ref.~\cite{lutz-korpa02}.

The $\Sigma(1385)$ moves very little in the medium, around 7 MeV  
from its free space position,
for both approximations. Again, in
the approximation that only dresses the antikaons this moderate
effect differs from the 60 MeV attraction quoted in
Ref.~\cite{lutz-korpa02}. The 30 MeV width of the $\Sigma(1385)$
in free space increases by about 10 MeV when antikaons are
dressed, due to the opening of new decay channels such as
$\Sigma^* N \to \Lambda N,\ \Sigma N$ through the coupling of the
${\bar K}$ to $Y N^{-1}$ excitations. For similar reasons mentioned above 
 in the case of the $\Sigma(1195)$, the decay width
into these new decay channels increases spectacularly when the
pions are also dressed, giving rise to a $\Sigma^*$ in the medium
having a width of around 80 MeV, consistently to what is found 
in Ref.~\cite{murat}, where the self-energy of the
$\Sigma(1385)$ is evaluated explicitly. Here, instead, we make 
a complete unitary theory of the $p$-wave $\bar{K}N$ scattering in the medium, 
where all particles and resonances are renormalized automatically
at the same time.

As one can see from Fig.~\ref{fig:sigma} we obtain an attractive $\Sigma$ self-energy in the medium  of around 35 MeV at $\rho$$=$$\rho_0$, much in line with what one obtains
from  \cite{Batty:1978sb} and \cite{pedro}.  In the literature one finds
potentials coming from fits to data of $\Sigma$-atoms with a chosen parametrization that
leads to repulsion at short distances \cite{Batty:1994yx}, while they are
attractive at large distances. A similar behaviour is obtained in models using
Dirac phenomenology \cite{Mares:1995bm}.  Other theoretical studies using
chiral perturbation theory find a net repulsion at long distances of the order
of 59 MeV at nuclear matter density \cite{Kaiser:2005tu}.  The situation is
thus confusing and the only experimental evidence is that atoms require an
attraction at the relatively large distances probed.  

There is another type of
experimental information provided by the study of the $\Sigma$-production
spectrum in the $(\pi^-, K^+)$ reactions. Analysis of the spectra within the
distorted wave approximation for the pion and kaon waves leads to a repulsive
$\Sigma$-nucleus potential \cite{Noumi:2001tx,Saha:2004ha,Kohno:2004pb}.  Yet,
here we must make some comment to clarify the origin of these results. The analyses of  \cite{Noumi:2001tx,Saha:2004ha,Kohno:2004pb} use distorted
waves for the pions and kaons or make use of the efficient and equivalent
Green's function method \cite{Morimatsu:1994sx}. This is appropriate when one
studies coherent production or elastic scattering, but not for inclusive
reactions as is the case there. The reason is simple: the optical potentials
used for pions and kaons have an imaginary part that comes from quasi-elastic
collisions and absorption.  This separation is done in potentials like 
the one given in \cite{Nieves:1991ye} for pions or in \cite{Oset:2000eg} for kaons. In the use of the
distorted waves the imaginary part of the potential depletes the strength of
the waves, in other words, every time there is a quasi-elastic collision or
absorption of the $K$ or $\pi$ the event is removed.  This is correct when one
looks at the formation of the ground state since any quasi-elastic collision
excites the nucleus. However,  in inclusive
reactions where one sums over all final nuclear states, this is not correct
because if there is a quasi-elastic collision the particle is still there and
can be detected, while the calculation has removed it. It is then clear that
the calculation will push to get a repulsive $\Sigma$-nucleus potential to 
prevent the $\Sigma$ from being too close to the nucleus where the
inappropriate calculation
of the distorted pion and kaon waves would remove too many events. 
The experimental
situation is thus confusing and the only firm information is the attractive
potential felt by the $\Sigma^-$ at the small densities probed by the atoms.

\begin{figure}
\begin{center}
    \includegraphics[width = \textwidth]{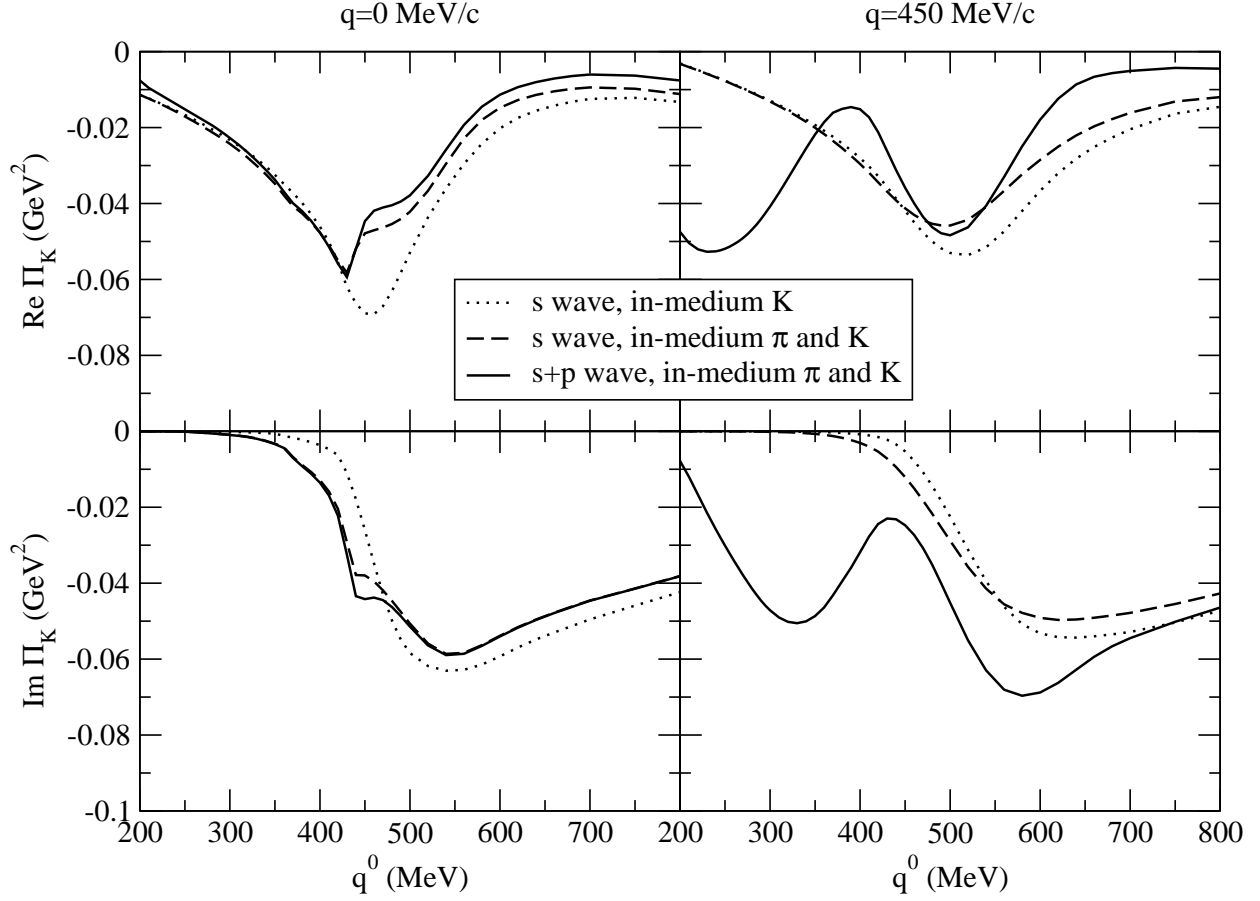}
\caption{Real part (upper panels) and imaginary part (lower
panels) of the antikaon self-energy as functions of the antikaon
energy $q^0$, for two values of the antikaon momentum, $q=0$ MeV/c
(left panels) and $q=450$ MeV/c (right panels). }
\label{fig:self-energy}
\end{center}
\end{figure}

The self-energy of the antikaon at $\rho=0.17$ fm$^{-3}$ as a function of
the antikaon energy is displayed in Fig.~\ref{fig:self-energy} for two
values of the antikaon momentum, $q=0$ and $450$ MeV/c. We show
the $s$-wave component of the self-energy for the approximation that
only dresses antikaons self-consistently (dotted line) and when the
dressing of pions is also incorporated (dashed lines). These
results correspond to those obtained in Ref.~\cite{Ramos:1999ku}.
The small amount of imaginary part at subthreshold antikaon energies and zero
momentum is due to $s$-wave excitations of the type
$\bar K N N \to \Sigma N,\ \Lambda N$, where the nucleons have assumed to
feel a potential of -70 MeV and the hyperons one of -30 MeV. 
Note that some extra enhancement of the
imaginary part would be visible in that region if we used a more moderate
attraction for the nucleon potential. 
The solid line shows the results when the new $p$-wave components
calculated in the present work are also incorporated. In the
case of zero antikaon momentum, the additional $p$-wave strength
corresponds to the $p$-wave induced $s$-wave corrections 
described in Sect.~\ref{sec:self} (see also 
Ref.~\cite{Garcia-Recio:1988bs}), which
produce a slight repulsion in the real part of the self-energy 
at the antikaon mass since the energies
that come into play are above $\Lambda$, $\Sigma$ and $\Sigma^*$
excitation. The role of the $p$-wave self-energy is more evident for
an antikaon momentum of 450 MeV/c. The imaginary part of the
corresponding self-energy clearly displays the signals of $\Sigma
N^{-1}$ and $\Sigma^* N^{-1}$ components around 300 MeV and 550
MeV, respectively. 
 More specifically, around 200 MeV below the
 antikaon mass, the width, $-2{\rm Im}\Pi_{\bar K}/(2q^0)$,
 increases considerably to about 160 MeV. At the same energy but for
 a more moderate momentum of 200
 MeV/c, this quantity would be divided by a factor $(450/200)^2\simeq 5$. 
 Should the antikaon mode achieve such an amount of attraction, these results
 show that the
 width of the bound state would be appreciable due to the $p$-wave
 components of the ${\bar K}$ self-energy.

\begin{figure}
\begin{center}
    \includegraphics[width = \textwidth]{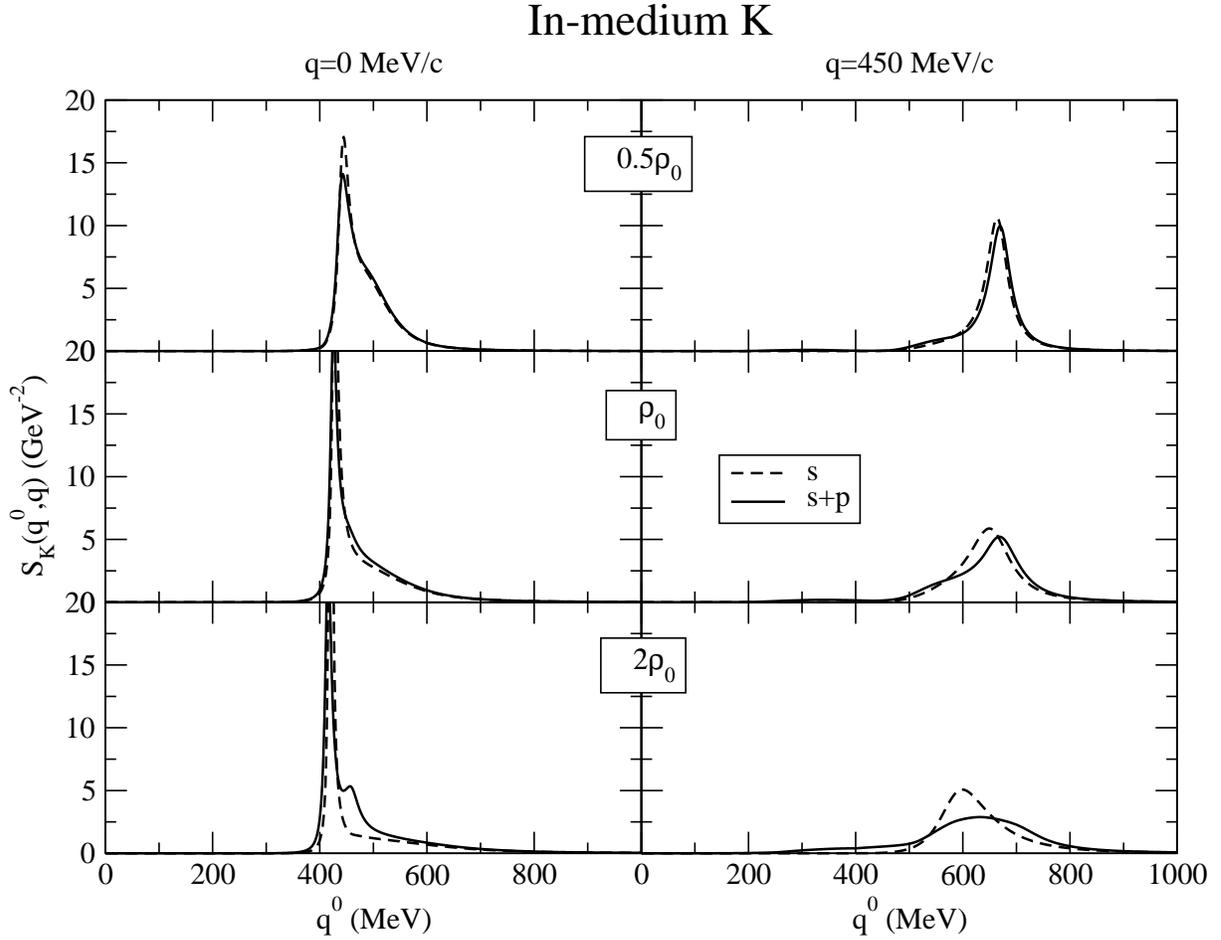}
\caption{Antikaon spectral function as a function of the antikaon energy
$q^0$, for two values
of the antikaon momentum, $q=0$ MeV/c (left panels) and $q=450$ MeV/c (right
panels), and three values of density, $\rho$$=$0.5$\rho_0$, $\rho_0$ and $2\rho_0$.
These results have been obtained for the approximation in which only the
antikaons are dressed self-consistently.}
\label{fig:spec.onlykaon}
\end{center}
\end{figure}

\begin{figure}
\begin{center}
    \includegraphics[width = \textwidth]{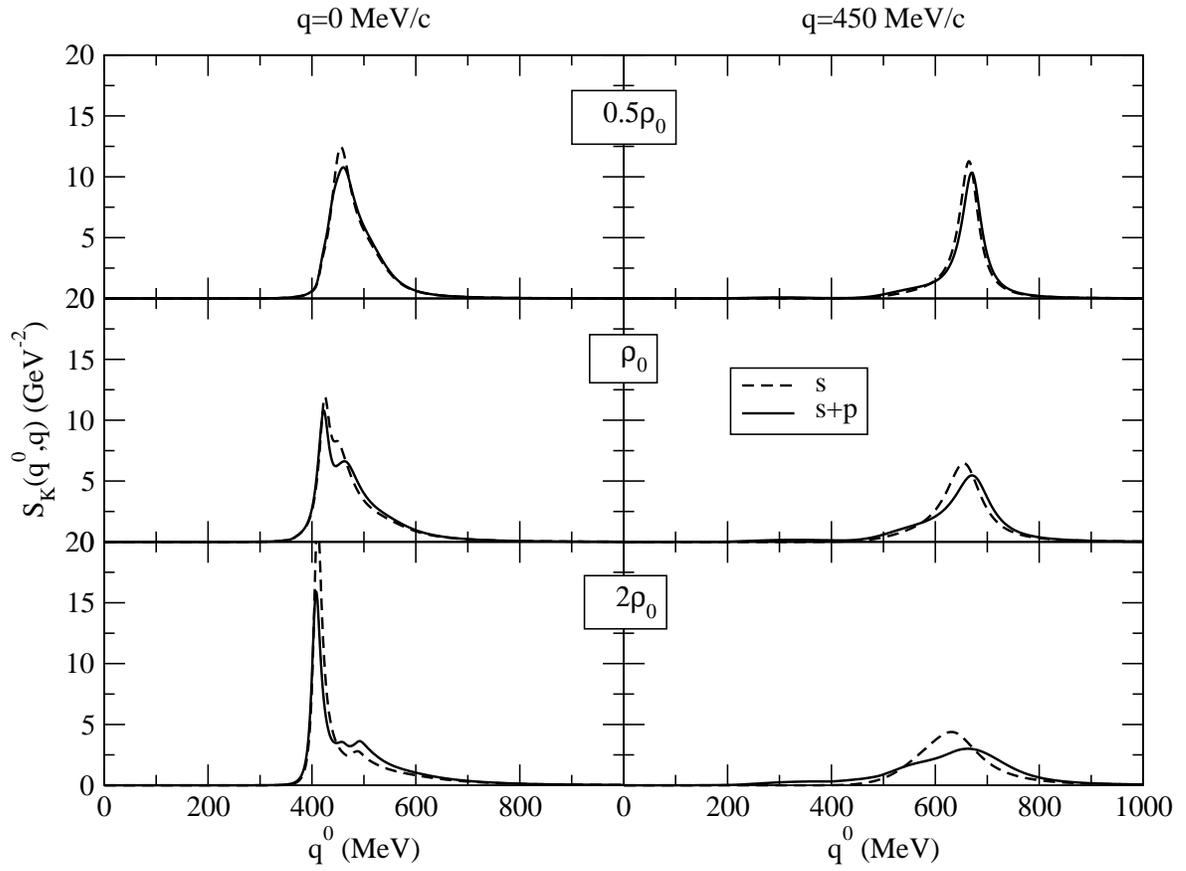}
\caption{The same as Fig.~\protect\ref{fig:spec.onlykaon} but with the
approximation that also considers the dressing of the pions.}
\label{fig:spec.pionkaon}
\end{center}
\end{figure}

We next comment our results on the antikaon spectral function for
the approximation that only dresses the antikaons in
Fig.~\ref{fig:spec.onlykaon} and when pions are also dressed in
Fig.~\ref{fig:spec.pionkaon}, for three different densities,
$\rho$$=$0.5$\rho_0$, $\rho_0$ and $2\rho_0$. As it is evident from
these plots, the antikaon spectral function is far from having a
Breit-Wigner type of behavior. At zero momentum, one observes the antikaon
quasiparticle peak, located at a lower energy than the position of the
free antikaon pole, superimposed to a shoulder of slow fall off on
the right-hand side, which corresponds
to $\Lambda(1405) N^{-1}$ excitation. At normal nuclear matter density, the
quasiparticle peak at zero momentum is displaced by 
about -60 MeV with respect to the free space position, while at a momentum of
450 MeV/c the displacement only amounts to about
-5 MeV.
We observe that, even at zero momentum, there
is a change in the $s$-wave spectral function (dashed lines)
when the $p$-wave strength is also included (solid lines), which is due to
the $p$-wave induced $s$-wave Fermi motion corrections. As density increases, the
quasiparticle peak at zero momentum gains attraction while the
spectral function at an antikaon momentum of 450 MeV/c gets
strongly diluted, especially when $p$-waves are incorporated. The
small peak in Fig.~\ref{fig:spec.onlykaon} to the right of 
the main quasiparticle peak for the $q=0$ spectral
function at $2\rho_0$ 
is due to the $p$-wave induced $s$-wave component associated to
$\Sigma^* N^{-1}$ excitations, which are located close to the dressed antikaon
quasiparticle peak. When pions
are also dressed the antikaon spectral functions displayed in
Fig.~\ref{fig:spec.pionkaon} show similar features, although
somewhat wider and more diluted.

\begin{figure}
\begin{center}
    \includegraphics[width = \textwidth]{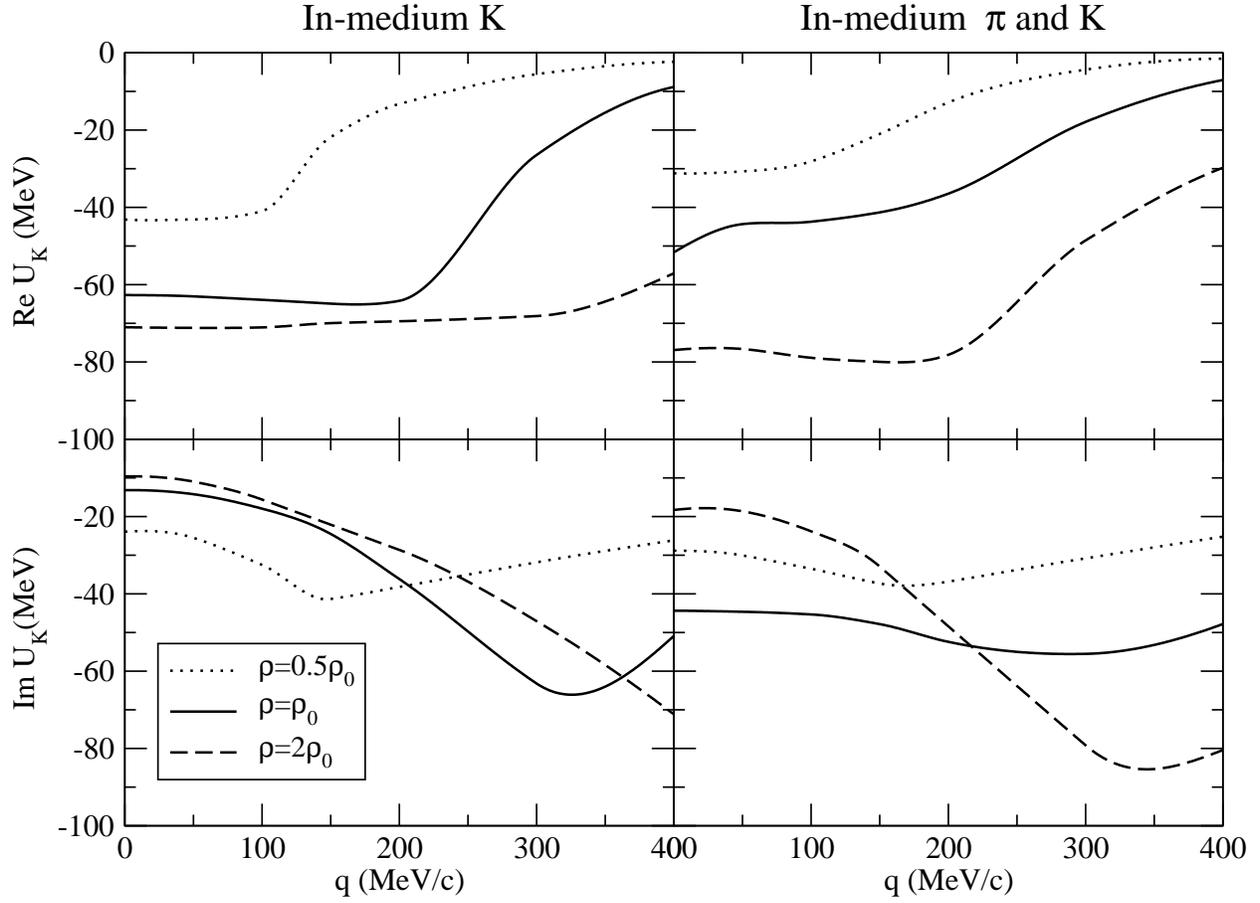}
\caption{Real part (upper panels) and imaginary part (lower panels) of the
antikaon optical potential as functions of the antikaon momentum for three
values of density, $\rho$$=$$0.5\rho_0$, $\rho_0$ and $2\rho_0$ and two
approximations, dressing only kaons (left panels) and dressing also the pions
(right panels).}
\label{fig:uk}
\end{center}
\end{figure}

We finally present in Fig.~\ref{fig:uk} results for the antikaon optical potential
calculated as $\Pi_k(q,\varepsilon(q))/(2\varepsilon(q))$, where
$\varepsilon(q)$ is the quasiparticle energy fulfilling
$\varepsilon(q)^2=m_K^2+q^2+{\rm Re}\Pi(q,\varepsilon(q))$, for three
different densities $\rho$$=$0.5$\rho_0$, $\rho_0$ and $2\rho_0$. The
real and imaginary parts of the antikaon optical potential as
functions of $q$ are shown on the left panels,
for the approximation that only dresses the
antikaons, and on the right panels, when pions are also dressed.
As density increases, the real part of the optical potential becomes more
attractive. In the approximation that only dresses the antikaons
self-consistently, the value of the optical potential at zero momentum goes from
around -40 MeV at $\rho_0/2$ to -70 MeV at $2 \rho_0$. The width of this zero
momentum state, which is twice the size of the corresponding imaginary part of the
optical potential, decreases with density due to the loss of decaying 
phase-space as the antikaon gains attraction.  This picture gets somewhat
distorted when pions are also dressed. Whereas, in this case,
the real part of the optical
potential at zero momentum goes from about -30 MeV at $\rho_0/2$ to almost
-80 MeV for $2\rho_0$, the size of the imaginary part first increases (from $\rho_0/2$
until $\rho_0$) and then decreases fast (from $\rho_0$ to $2\rho_0$).
This is because part of the loss of decaying phase space with
increasing density is compensated by the appearance of
new decaying states, $Y N N^{-1}$, the amount of which increases with increasing
density.  
Our results are qualitatively very similar to those shown in Ref.~\cite{laura},
where a self-consistent calculation of the antikaon optical potential using the
meson-exchange J\"ulich ${\bar K}N$ interaction was presented.

\section{Conclusions}
\label{conclusions}

We have investigated the properties of the ${\bar K}$ self-energy in nuclear
matter after incorporating the medium modifications on the $s$-wave and 
$p$-wave ${\bar K}N$ amplitudes, within the context of a chiral unitary
approach. The $s$-wave interaction is taken from the Weinberg-Tomozawa term and
the $p$-wave collects a small contribution from this source and a large
contribution from the $\Lambda$, $\Sigma$ and $\Sigma(1385)$ pole terms. To
account for the medium renormalization of the amplitudes, 
we include, in a
self-consistent way, Pauli blocking effects, meson self-energies corrected 
by nuclear short-range correlations and baryon binding potentials.

We have payed a special attention to the modification of the $p$-waves,
showing that for the in-medium corrections it is not possible to apply the
on-shell factorization of the amplitudes, which is the standard procedure in
free space.

The $\Lambda$ and $\Sigma$ in nuclear matter at saturation density feel an
attractive potential of around -30 MeV, while the $\Sigma^*$ stays pretty much
at its free space position but its width is sensibly increased to
about 80 MeV. 

The ${\bar K}$ self-energy is evaluated as a function of laboratory
 energy $q^0$ and momentum $\vec{q}$, which are independent variables
 in the medium. The $p$-wave contributions to the antikaon self-energy are small for
 low momentum kaons. However, at large momentum values of about 450 MeV/c, 
 one finds considerable strength at subthreshold energies coming from ${\bar K}N \to
 \Sigma$ conversion. 
 
 The ${\bar K}$ spectral function shows a distinct quasiparticle peak around 60
 MeV below the antikaon mass from zero momentum. The peak moves to higher energy
 and gets closer to the free space position as the ${\bar K}$ picks up momentum.
 Thus, the consideration of $p$-waves does not help in increasing the binding
 energy of the ${\bar K}$ mode. The increased amount of strength at low 
 antikaon energies of around 300 MeV when $p$-waves are included implies that, if the 
 kaon mode was able to generate such an amount of attraction, the bound state
 would have an appreciable width. However, the antikaon optical
 potential obtained in the present work can only give ${\bar K}$ states in 
 matter bound by no more than 50 MeV and having a width of the 
 order of 100 MeV, in qualitative agreement with all existing
 self-consistent calculations.

\section*{Acknowledgments}
L.T. wishes to acknowledge support from Gesellschaft f\"ur Schwerionenforschung
and  Alexander von Humboldt Foundation.  This work is partly supported by
contracts BFM2003-00856 and FIS2005-03142 from
MEC (Spain) and FEDER,
the Generalitat de Catalunya contract 2005SGR-00343,
and the E.U. EURIDICE network contract HPRN-CT-2002-00311.
This research is part of the EU Integrated Infrastructure Initiative
Hadron Physics Project under contract number RII3-CT-2004-506078.

\end{document}